\documentclass[journal]{IEEEtran}
\usepackage{amsmath,amssymb,amsthm} 
\usepackage{mathrsfs}
\usepackage{algorithm}  
\usepackage{algpseudocode}

\usepackage{graphicx}
\graphicspath{{figures/}}

\makeatletter
\@ifclasswith{IEEEtran}{compsoc}{%
	\usepackage[caption=false,font=normalsize]{subfig} 
}{%
	\usepackage[caption=false,font=footnotesize]{subfig} 
}

\usepackage{etoolbox}    

\usepackage{stfloats}
\AtBeginDocument{
	\addtolength{\abovedisplayskip}{-1ex}
	\addtolength{\belowdisplayskip}{-1ex}
	\addtolength{\abovedisplayshortskip}{-1ex}
	\addtolength{\belowdisplayshortskip}{-1ex}
}

{\tiny }

\usepackage{enumitem}
\setlist{nosep,left=0pt}

\makeatletter
\patchcmd{\algorithmic}{\itemsep\z@}{\itemsep2pt}{}{}
\makeatother

\usepackage{xcolor}  
\usepackage{bm}
\usepackage{cite}
\usepackage[colorlinks,
urlcolor=blue,
linkcolor=blue,
anchorcolor=blue,
citecolor=blue]{hyperref}

\usepackage{lettrine}

\usepackage{array}       
\usepackage{tabularx}    
\usepackage{booktabs}    
\usepackage{multirow}    

\newcolumntype{Y}{>{\raggedright\arraybackslash}X}

\begin{document}
	
	\title{Design of Multi-UAV Cooperative Deep Semantic Autoencoders for Communication Networks}

	\author{
		Xiaoling Han,~\IEEEmembership{Student Member,~IEEE,}
		Bin Lin*,~\IEEEmembership{Senior Member,~IEEE,}
		Nan Wu,~\IEEEmembership{Member,~IEEE,}
		Tony Q. S. Quek,~\IEEEmembership{Fellow,~IEEE}
		
		
		\thanks{
			The work was supported by the National Natural Science Foundation of China (No. 62371085) and Fundamental Research Funds for the Central Universities (No. 3132023514). This paper was supported in part by the National Research Foundation, Singapore and Infocomm Media Development Authority under its Future Communications Research \& Development Programme. This paper was supported by the China Scholarship Council (CSC).
			(Corresponding author: Bin Lin.)
		}
		
		\thanks{
			Xiaoling Han, Bin Lin, Nan Wu are with the Department of Information Science and Technology of Dalian Maritime University, Dalian, China, 116026 (email: xiaolinghan@dlmu.edu.cn, binlin@dlmu.edu.cn, wu.nan@dlmu.edu.cn). 
			Tony Q. S. Quek is with the Information Systems Technology and Design Pillar, Singapore University of Technology and Design, Singapore, 487372 (email: tonyquek@sutd.edu.sg).
		}
		
		\thanks{
			This manuscript has been accepted by IEEE Transactions on Vehicular Technology, DOI: 10.1109/TVT.2026.3673755.
		}
	}
	
	\markboth{Journal of \LaTeX\ Class Files,~Vol.~14, No.~8, August~2015}%
	{Shell \MakeLowercase{\textit{\textit{et al.}}} Bare Demo of IEEEtran.cls for IEEE Journals}
	
	\maketitle
	
	\begin{abstract}
		Driven by the rapid development of the low-altitude economy, Uncrewed Aerial Vehicle (UAV) networks have become increasingly important for executing tasks such as search, rescue, and monitoring.
		However, under bandwidth-constrained scenarios, UAV networks still face significant challenges in task execution efficiency and intelligent processing capabilities. To address these issues, we propose Multi-UAV Cooperative Deep Semantic Autoencoders (MC-DSA) for communication networks. 
		The architecture comprises one Task UAV (T-UAV) and multiple Support UAVs (S-UAVs), which jointly enhance communication efficiency and data-processing capability via semantic communication and cooperation. Specifically, we introduce a semantic matching and source selection mechanism, enabling the proposed MC-DSA to handle low-quality images while ensuring effective utilization of high-quality images. The T-UAV broadcasts a task query, and each S-UAV computes the semantic similarity between its local image and the query. Communication links are established only when the similarity exceeds a predefined threshold. The training algorithm optimizes network parameters through gradient descent, improving semantic feature extraction accuracy. The testing algorithm evaluates UAV collaboration efficiency and communication quality. Experimental results demonstrate that, compared with benchmarks, the proposed MC-DSA improves image semantic segmentation accuracy while maintaining task execution efficiency. As a result, it reduces redundant transmissions and offers an effective solution for multi-UAV communication networks.
	\end{abstract}
	
	\begin{IEEEkeywords}
		UAV collaboration, source selection, semantic communication, deep learning
	\end{IEEEkeywords}
	
	\IEEEpeerreviewmaketitle
	\section{Introduction}
	
	\lettrine[lines=2, loversize=0.1]{I}{n} Internet of Things (IoT) deployments, reliable scene understanding relies on intelligent communication \cite{IEEE6GEdge}. Conventional communication frameworks based on Shannon's information theory, which separate the physical layer from the application layer, fail to efficiently process high-dimensional semantic data \cite{zhishiku}. In multi-Uncrewed Aerial Vehicle (UAV) collaboration, semantic-aware networking confronts inconsistent semantic representations, imperfect cooperative communication mechanisms, and intensified resource competition \cite{yuyiganzhi}. These issues jointly manifest as redundant uplink transmissions and weak interference resilience under bandwidth-constrained conditions, ultimately limiting task execution efficiency \cite{maritimeUAV}. A mismatch exists between syntactic-level transmission fidelity and downstream task utility in multi-UAV collaboration. This motivates the task-oriented communication that reasons about the end task rather than the symbol stream.
	
	Semantic communication, as a task-oriented communication paradigm, has attracted extensive attention \cite{tiaozhan}. 
	By integrating Deep Learning (DL) technologies, it achieves a significant leap in data transmission from the traditional syntactic level to the semantic level \cite{engineeringsemantic}. 
	Unlike traditional communication, which is fundamentally based on signal transmission at the syntactic level, semantic communication pays more attention to semantic understanding and intelligent processing \cite{yuyichaoyuebite}. When processing text, images, and speech, semantic communication enables efficient information transmission and semantic-level inference \cite {engineeringsemantic, yuyituxiang1, crossmodal}. However, most existing semantic communication systems are designed for non-cooperative point-to-point links, which limits their applicability in multi-user and cooperative settings.
	
	Faced with complex tasks, a single UAV often struggles to obtain comprehensive data independently due to perspective limitations and low-altitude airspace constraints. Multi-UAV cooperation compensates for these limitations through cooperative sensing and real-time information sharing. This coordination enables more reliable perception and decision-making. Recent studies in multi-user semantic communication have demonstrated that collaboration across users can improve task execution by facilitating information interaction. A multi-user communication system jointly optimizes the semantic encoder and decoder to maximize task-level performance \cite {liu2020when2com}.
	Xie \textit{et al.} \cite {multiuser2} constructed a task-oriented multi-user semantic communication system. By jointly optimizing the transmitter and receiver, they captured the features of multimodal data and achieved efficient transmission for visual question answering tasks. 
	Wei \textit{et al.} \cite {multiuser4} proposed a deep over-the-air computation based multi-user semantic communication system. It co-designs communication and computation, achieves multi-user information fusion, and fine-tunes both the semantic and computing modules, leading to stable performance in complex smart-city tasks.
	Chen \textit{et al.} \cite {multiuser3} designed a novel pilot-free semantic communication system for image transmission over multi-user multiple-input multiple-output fading channels. By extracting multi-scale semantic features from the source images, they reconstructed the images at the receiver and introduced the ResNet transformer module to further enhance the performance.
	
	Furthermore, for semantic communication, image segmentation is a particularly important downstream task \cite{mIOU}. Image segmentation converts the original pixels into dense pixel-by-pixel semantics, thereby directly driving functional tasks such as navigation, path planning, collision avoidance, and search and rescue \cite{segmentpaper}. Compared with classification, segmentation exposes the complete spatial structure of the scene. In multi-user scenarios, effective collaboration hinges on cross-user semantic alignment and on-demand decisions, such as when to communicate, who should transmit, and what to transmit. These call for selective mechanisms that make the collaboration policy prioritize task utility.

	For practical multi-user deployments, it is also essential to balance task effectiveness against communication and computation resource consumption. Many studies have investigated this problem. Zeng \textit{et al.} \cite {multiuser5} proposed the diffusion model channel enhancer for multi-user semantic communication systems. Aiming to address the impact of wireless channels, they optimized the transmission and fusion processes of multi-user semantic features. The channel state estimation improved the recovery effect of semantic features at the receiver. 
	Huang \textit{et al.} \cite {multiuser1} proposed the DeepSC for multi-user efficient transmission. By using dedicated flag vectors to identify users and embedding them at both the transmitter and receiver, they compressed multi-user semantic messages through a fusion mechanism. This enables concurrent message transmission and reduces the transmission burden.
	Xing \textit{et al.} \cite {multiuser6} constructed a deep semantic communication system based on federated learning with dynamic model aggregation, which can adapt to time-varying channels. By adjusting the model aggregation weights, it balances the effectiveness and diversity of the models, thus optimizing the performance of multi-user semantic communication. 
	Multi-UAV collaboration cannot overlook lightweight design and real-time collaboration needs. Thus, a cooperative mechanism is required to streamline task execution and reduce resource overhead.
	
	Based on the above analysis, we propose Multi-UAV Cooperative Deep Semantic Autoencoders (MC-DSA) for communication networks. 
	It targets task-oriented collaboration in UAV swarms under resource constraints. The proposed MC-DSA uses semantic matching to activate selective transmission, enabling collaboration among UAVs while accounting for communication load. The main contributions of this paper are as follows.
	
	\begin{enumerate}
		
		\item Novel semantic communication solution: we propose MC-DSA, a cooperative semantic communication framework for multi-UAV networks, consisting of one Task UAV (T-UAV) and multiple Support UAVs (S-UAVs). By transmitting compact semantic features, the framework aligns communication with task relevance and improves efficiency under bandwidth, latency, and energy constraints.

		\item New collaboration mechanism: we design a policy to determine when and who should communicate. Each S-UAV computes the semantic similarity between its local observation and the T-UAV’s task query. S-UAVs whose scores exceed a threshold are admitted to transmit. This mechanism forms communication groups, suppresses redundant uploads, and realizes implicit quality-aware collaboration learned directly from task supervision.
		
		\item New training and testing algorithms: we develop a centralized training algorithm with soft selection and an adaptive learning-rate schedule, enabling stable convergence. All modules are optimized with validation monitoring. We further design a testing algorithm with hard gating to evaluate task accuracy. This design maintains consistency between optimization and deployment.
		
		\item Comprehensive experimental validation: we assess the proposed MC-DSA under diverse channel conditions and different experimental settings. Results show that, compared to benchmarks, the proposed MC-DSA attains higher task performance while reducing communication resource usage. It maintains stable performance and converges smoothly, demonstrating robustness, scalability, and resource efficiency for multi-UAV communications.

	\end{enumerate}
	
	The remainder of this paper is organized as follows. Section II presents the proposed MC-DSA, including the communication model and resource model. Section III describes the training and testing strategy. Section IV presents the performance evaluation under different settings. Section V concludes the paper and discusses future work.
	
	\section{System Model}
	
	\begin{figure*}[t]
		\centering
		\includegraphics[width=1\linewidth]{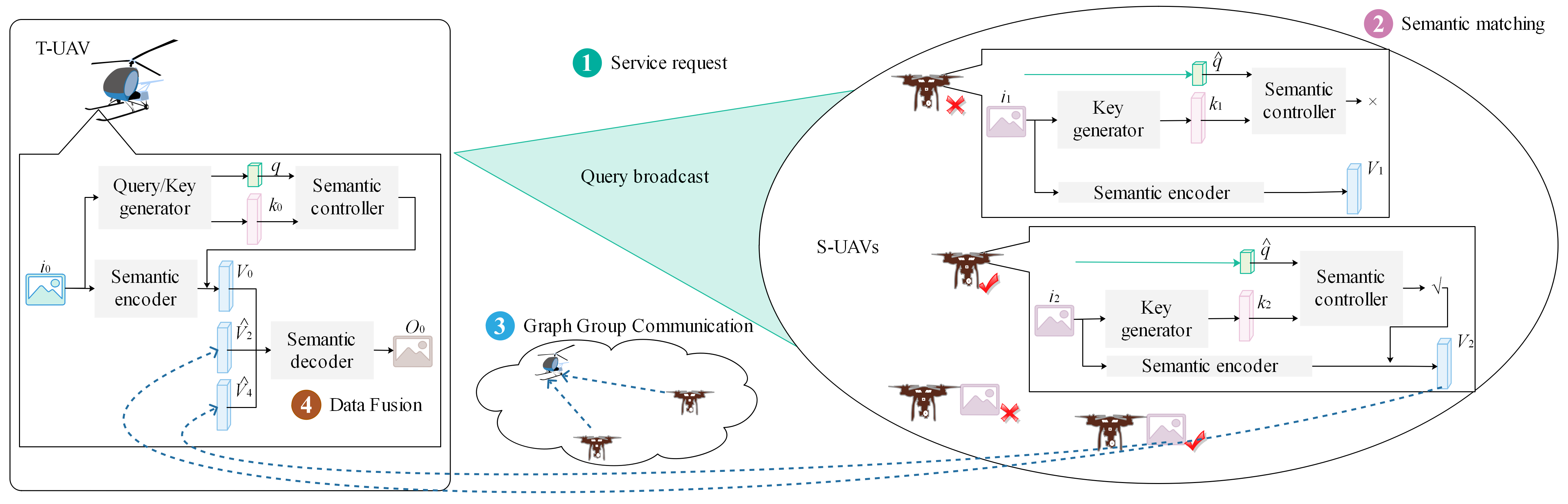}
		\caption{The system model of the proposed MC-DSA.}
		\label{systemmodel}
	\end{figure*}

	In this section, we establish the system model of MC-DSA and define the key components, including the communication and resource models as well as the optimization objective.
	
	The proposed MC-DSA is shown in Fig. \ref{systemmodel}. It consists of one T-UAV, denoted $U_0$, and $N$ S-UAVs, collected as $\mathcal{U}=\{U_1,\dots,U_N\}$. At each frame, $U_i$ acquires one RGB image $\bm{i}_i\in\mathbb{R}^{C_{\rm in}\times H\times W}$. The downstream task is semantic segmentation with $n_{\rm class}$ classes, which is executed at the T-UAV. Uplink access follows Time Division Multiple Access (TDMA). Since this paper focuses on the physical and link layers of multi-UAV cooperative communication, the UAV flight trajectories for image acquisition are fixed and known.
	The network architecture is summarized in Tab. \ref{tab_arch_when2com}. $B$ is the batch size and $H\times W$ denotes the spatial resolution. $L_{\rm q}$ and $L_{\rm k}$ denote the query and key dimensions, respectively.

	\begin{table*}[t]
		\centering
		\caption{Network architecture of the proposed MC-DSA.}
		\label{tab_arch_when2com}
		\renewcommand{\arraystretch}{1.05}
		\begin{tabularx}{\textwidth}{@{}p{3cm} p{3cm} p{4cm} l@{}}
			\toprule
			\textbf{Category} & \textbf{Module} &\textbf{Layer Name} & \textbf{Output dimensions} \\
			\midrule
			\multirow{16}{*}{Semantic encoder}
			&& Input & $[B,N,C_{\rm{in}},H,W]$ \\

			&& Conv blocks & $ [B \cdot N,C_{\mathrm{feat}}/8,H/2,W/2]$ \\
			
			&& Conv blocks & $ [B \cdot N,C_{\mathrm{feat}}/4,H/4,W/4]$ \\

			&& Conv blocks & $ [B \cdot N,C_{\mathrm{feat}}/2,H/8,W/8]$ \\
			
			&& Conv blocks & $ [B \cdot N,C_{\mathrm{feat}},H/16,W/16]$ \\
			
			&& Conv2d + BatchNorm2d & $ [B \cdot N,C_{\mathrm{feat}}/4,H/16,W/16]$ \\

%

			&& Conv2d & $ [B \cdot N,C_{\mathrm{feat}}/4,H/16,W/16]$ \\
			&& Patch embed & $ [B \cdot N,H/16\times W/16,C_{\mathrm{feat}}/4]$ \\
			&& RSTB & $ [B \cdot N,H/16\times W/16,C_{\mathrm{feat}}/4]$ \\
			&& RSTB & $ [B \cdot N,H/16\times W/16,C_{\mathrm{feat}}/4]$ \\
			&& RSTB & $ [B \cdot N,H/16\times W/16,C_{\mathrm{feat}}/4]$ \\
			&& RSTB & $ [B \cdot N,H/16\times W/16,C_{\mathrm{feat}}/4]$ \\
			&& Patch unembed & $ [B \cdot N,C_{\mathrm{feat}}/4,H/16,W/16]$ \\
			&& Conv2d &$ [B \cdot N,C_{\mathrm{feat}}/4,H/16,W/16]$ \\
			&& Conv2d & $ [B \cdot N,C_{\mathrm{feat}}/4,H/16,W/16]$ \\

			&& Conv2d + BatchNorm2d & $ [B \cdot N,C_{\mathrm{feat}},H/16,W/16]$ \\
			
			&& Conv blocks & $ [B \cdot N,C_{\mathrm{feat}},H/32,W/32]$ \\

			\midrule
			
			\multirow{15}{*}{Q/K generator}
			& \multirow{4}{*}{Policy encoder}
			
			& Input & $[B,N,C_{\rm{in}},H,W]$ \\
			
			&& Same as semantic encoder  & $ [B \cdot N,C_{\mathrm{feat}},H/32,W/32]$ \\
			

			&& Conv blocks & $ [B \cdot N,C_{\mathrm{feat}}/2,H/64,W/64]$ \\

			&& Conv blocks & $ [B \cdot N,C_{\mathrm{feat}}/2,H/128,W/128]$ \\
			
			\cline{2-4}
			&\multirow{7}{*}{Key net}
			& Pool & $ [B \cdot N,C_{\mathrm{feat}}/2,P,P]$ \\
			&& LayerNorm & $[B \cdot N,C_{\mathrm{feat}}/2\times P\times P] $ \\
			&& Linear & $ [B \cdot N,C_{\mathrm{feat}}/2]$ \\
			&& ReLU & $ [B \cdot N,C_{\mathrm{feat}}/2]$ \\
			&& Linear & $ [B \cdot N,C_{\mathrm{feat}}/4]$ \\
			&& ReLU & $ [B \cdot N,C_{\mathrm{feat}}/4]$ \\
			&& Linear & $ [B \cdot N,L_{\rm{k}}]$ \\
			
			\cline{2-4}
			&\multirow{7}{*}{Query net}
			& Pool & $ [B \cdot N,C_{\mathrm{feat}}/2,P,P]$ \\
			&& LayerNorm & $ [B \cdot N,C_{\mathrm{feat}}/2\times P\times P]$ \\
			&& Linear & $[B \cdot N,C_{\mathrm{feat}}/2]$ \\
			&& ReLU & $ [B \cdot N,C_{\mathrm{feat}}/2]$ \\
			&& Linear & $ [B \cdot N,C_{\mathrm{feat}}/4]$ \\
			&& ReLU & $ [B \cdot N,C_{\mathrm{feat}}/4]$ \\
			&& Linear & $ [B \cdot N,L_{\rm{q}}]$ \\
			
			\midrule
			
			\multirow{2}{*}{Semantic controller}
			&& Linear & $[B,1,L_{\rm{q}}] \rightarrow [B,1,L_{\rm{k}}]$ \\
			&& Softmax & $ [B,N,1]$ \\
			\midrule
			
			\multirow{4}{*}{Semantic decoder}
			&& Conv2d & $ [B,C_{\mathrm{feat}}/2,H/32,W/32]$ \\
			&& ReLU & $ [B,C_{\mathrm{feat}}/2,H/32,W/32]$ \\
			&& Conv2d & $ [B,n_{\rm{class}},H/32,W/32]$ \\
			&& Upsample & $ [B,n_{\rm{class}},H,W]$ \\
			\bottomrule
		\end{tabularx}
	\end{table*}
	
\begin{figure}[t]
		\centering
		\subfloat[Fully connected\label{modelA}]{
			\includegraphics[width=0.48\linewidth]{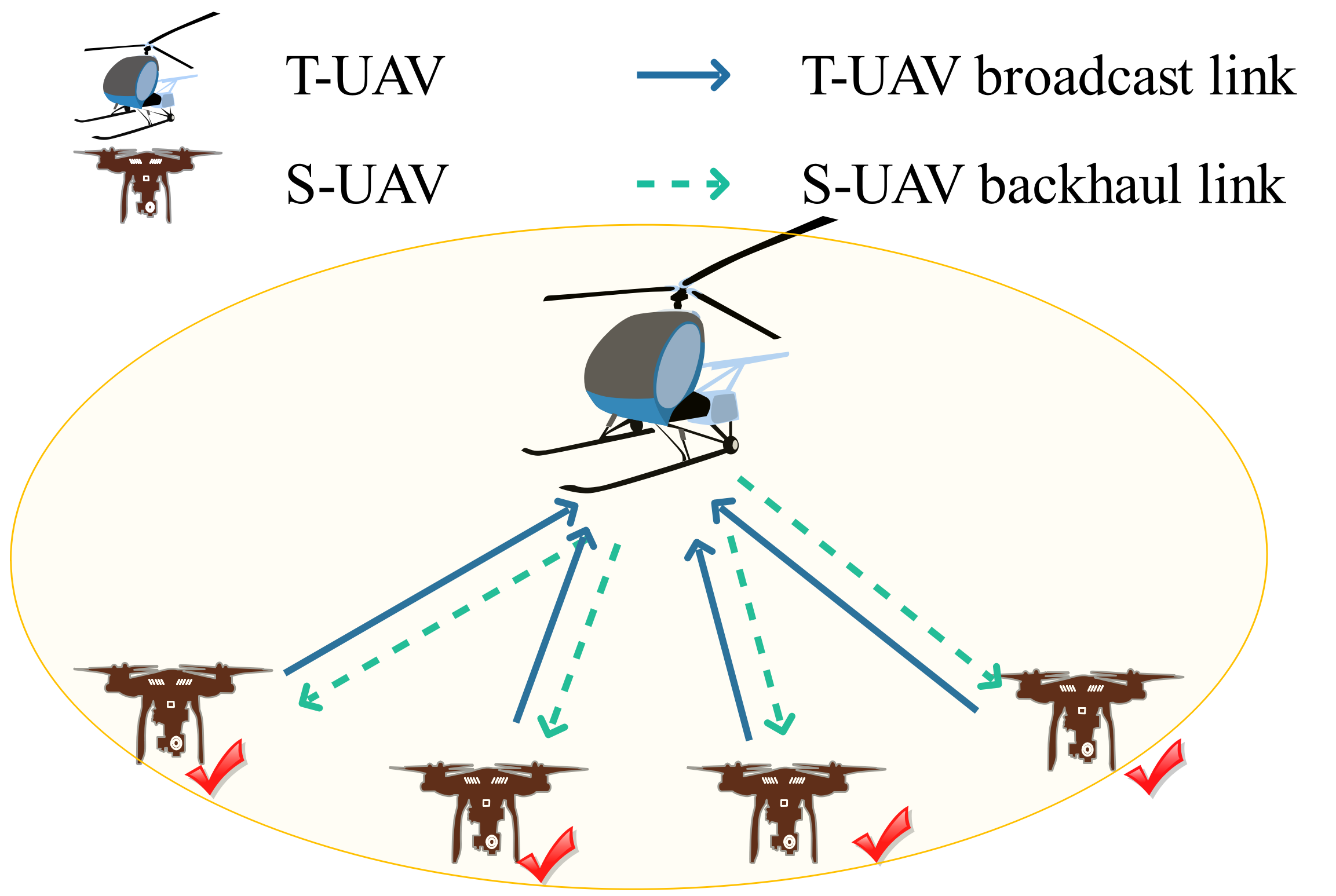}}\hfill
		\subfloat[Graph group\label{modelB}]{
			\includegraphics[width=0.48\linewidth]{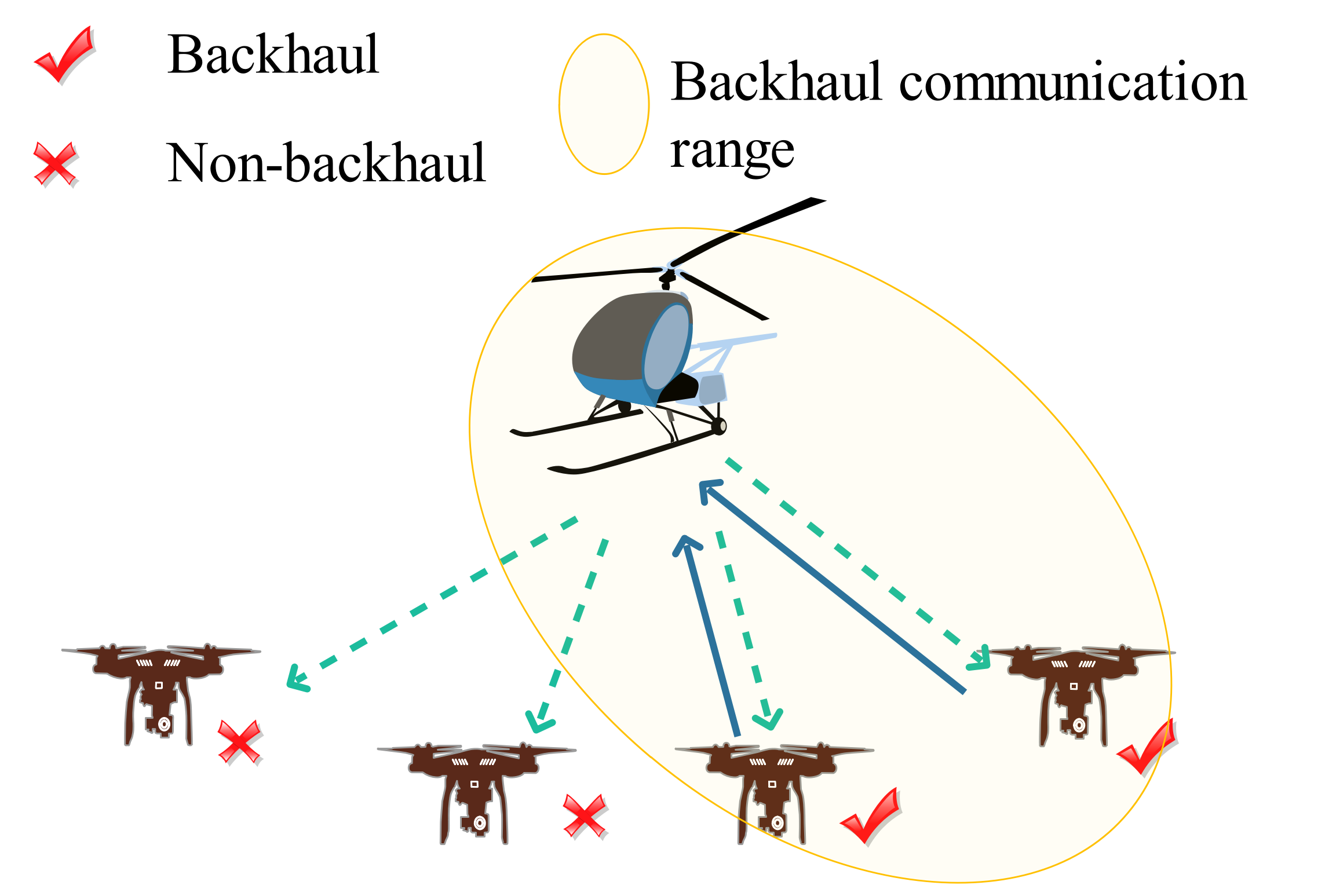}}
		\caption{Different cooperative communication models.}
	\end{figure}

	The T-UAV is responsible for directly executing tasks. S-UAVs provide complementary observations to assist task completion. Conventional multi-UAV schemes typically assume a fully connected all-to-one uplink. When the T-UAV lacks sufficient information, all S-UAVs transmit their data to the T-UAV indiscriminately as shown in Fig. \ref{modelA}, resulting in unnecessary bandwidth consumption. The proposed MC-DSA adopts a graph-based group communication paradigm \cite{liu2020when2com} as shown in Fig. \ref{modelB}. Task success depends on the information quality of the T-UAV’s observation. When the T-UAV image does not meet the quality requirement, the T-UAV requests collaboration to improve task performance. Semantic matching drives the T-UAV to form a sparse communication group that contains the T-UAV and the S-UAVs with high task relevance. In bandwidth-constrained scenarios, the communication group is recomputed every frame, so the topology adapts to task demands and scene dynamics, reducing redundant transmissions and improving bandwidth utilization.
	
	Each UAV is equipped with two coordinated branches: (i) a task branch that extracts semantic features for the downstream image segmentation task and decides what content to transmit, and (ii) a policy branch that decides when to communicate (when2com) and with whom to communicate (who2com).
	In the task branch, a semantic encoder transforms the captured image into a compact, task-oriented representation that emphasizes object boundaries, region consistency, and long-range context. A lightweight semantic decoder maps fused features back to pixel-level logits for segmentation. The proposed MC-DSA places most capacity in the semantic encoder and keeps the semantic decoder shallow. This design recovers pixel-level logits with minimal computation and reduces latency and memory without degrading segmentation accuracy.
	In the policy branch, the query generator consists of a policy encoder followed by a query net, producing a low-dimensional query for control-channel broadcast. The key generator likewise consists of a policy encoder followed by a key net, producing higher-dimensional keys for robust matching. A semantic controller matches the query to keys, forms the communication group based on semantic similarity.

	\subsection{Communication Model}
	
	At every frame, the T-UAV captures $\bm{i}_0$ and each S-UAV $U_i$ captures $\bm{i}_i$. All UAVs use the same semantic encoder architecture and share a single parameter set, which ensures feature consistency and facilitates reliable cross-UAV matching. The semantic encoder follows a hierarchical feature-extraction architecture. 
	It consists of multiple Convolutional (Conv) blocks, each composed of a convolution layer, batch normalization, and ReLU activation, which progressively capture local spatial structures such as edges, textures, and object boundaries. 
    To enhance global representation, we incorporate the Swin Transformer, implemented as Residual Swin Transformer Blocks (RSTBs), to capture long-range dependencies and emphasize task-relevant semantic regions for the downstream task \cite{liu2021swin}.
	This design combines the local-detail sensitivity of convolution layers with the contextual reasoning ability of transformers, generating task-oriented semantic features. The semantic feature maps produced by all UAVs can be expressed as
	\begin{equation}
		\bm{V}_i \;=\; f_{\mathrm{en}}\!\big(\bm{i}_i, \bm{\epsilon}_{\mathrm{en}}\big)\ \in\ \mathbb{R}^{C_{\mathrm{feat}}\times (H/32)\times(W/32)},
	\end{equation}
	where $f_{\mathrm{en}}(\cdot,\cdot)$ and $\bm{\epsilon}_{\mathrm{en}}$ denote the mapping function and parameters of the semantic encoder. The feature tensor has $C_{\mathrm{feat}}$ channels and spatial resolution $(H/32)\times(W/32)$. The communication process is described below.

	\subsubsection{Service Request}
	
	The T-UAV generates and broadcasts a query signal $\bm{q}$ via a query generator, which is composed of a policy encoder and a query net. The policy encoder shares the same architecture as the semantic encoder but uses separate parameters, enabling it to learn the communication policy without affecting task-feature learning. Given the local input $\bm{i}_0$, it produces a policy feature, given by
	\begin{equation}
		\bm{V}_{\rm qk}^0 \;=\; f_{\mathrm{pen}}(\bm{i}_0, \bm{\epsilon}_{\mathrm{pen}}) \;\in\; \mathbb{R}^{C_{\mathrm{feat}}\times (H/32)\times(W/32)} .
	\end{equation}
	To obtain a fixed-length representation independent of the input resolution, the query net first applies an adaptive average pooling operator $p$ that maps any $H\times W$ feature to a compact $P\times P$ grid while retaining spatial layout, given by
	\begin{equation}
		\bm{z}_{\rm q}=\mathrm{vec}\!\big(p(\bm{V}_{\rm qk}^0)\big).
	\end{equation}
	A light Multilayer Perceptron (MLP) projects $\bm{z}_{\rm q}$ to an ultra-short vector, given by
	\begin{equation}
		\bm{q}=f_{\rm q}(\bm{z}_{\rm q};\bm{\epsilon}_{\rm q})\in\mathbb{R}^{L_{\rm q}},
	\end{equation}
	where $\bm{\epsilon}_{\mathrm{q}}$ denotes the learnable parameters of the query net. The query $\bm{q}$ is broadcast by the T-UAV to all S-UAVs over the wireless channel, modeled by multiplicative fading and additive noise, given by
	\begin{equation}
		\widehat{\bm{q}} \;=\; \bm{h}_i \bm{q} \;+\; \bm{n}_i ,
		\label{eq:uplink-q}
	\end{equation}
	where $\bm{h}_i$ denotes the fading gain within the frame, and $\bm{n}_i$ is additive noise with i.i.d. zero-mean entries and variance $\sigma_n^2$. This resolution-agnostic pooling makes the query representation stable across input sizes and encoder strides, preserves coarse spatial cues that benefit matching, and avoids hand-tuning for different image resolutions. The payload is compact and carries only task intent rather than raw features. The query is used only within the current frame.
	
	\subsubsection{Semantic Matching and Source Selection}
	
	Each UAV produces a key that summarizes its current observation for matching. For S-UAV $U_i$, the key $\bm{k}_i$ is generated by a key generator, which consists of a policy encoder and a key net. The policy encoder has the same architecture as the T-UAV policy encoder. The policy feature can be expressed as
	\begin{equation}
		\bm{V}_{\rm qk}^i \;=\; f_{\mathrm{pen}}(\bm{i}_i, \bm{\epsilon}_{\mathrm{pen}}) \;\in\; \mathbb{R}^{C_{\mathrm{feat}}\times (H/32)\times(W/32)} .
	\end{equation}
	Analogous to the query net, the key net first applies adaptive pooling to obtain a fixed-length summary of $\bm{V}_{\rm qk}^i$ and then uses an MLP to project it, but to a higher-dimensional space.
	The key can be expressed as
	\begin{equation}
		\bm{k}_i \;=\; f_{\rm k}(\bm{V}_{\rm qk}^i,\bm{\epsilon}_{\rm k}) \in \mathbb{R}^{L_{\rm k}},\qquad L_{\rm q}\ll L_{\rm k},
	\end{equation}
	where $\bm{\epsilon}_{\mathrm{k}}$ denotes the learnable parameters of the key net.
	The T-UAV also computes a self-key $\bm{k}_0$ from $\bm{V}_{\rm qk}^0$ to assess whether local data alone suffices. Each key is cached locally for one frame and is released before the next frame.
	
\begin{figure}[t]
		\centering
		\includegraphics[width=1\linewidth]{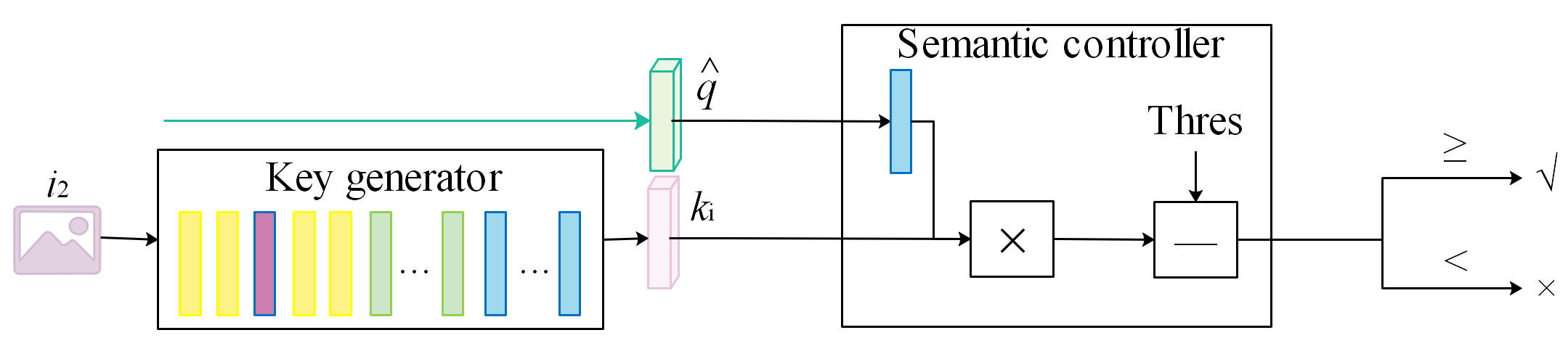}
		\caption{The workflow of semantic matching.}
		\label{simcompare}
	\end{figure}

	Next, the semantic controller computes pairwise similarities between the received query  $\widehat{\bm{q}}$ and each key $\bm{k}_i$ to quantify source importance, as shown in Fig.~\ref{simcompare}. A low-dimensional $\widehat{\bm{q}}$ minimizes control-channel overhead and interference risk, while higher-dimensional $\bm{k}_i$ preserves discriminative context for robust matching under viewpoint changes. The query is projected into the key space, given by
	\begin{equation}
		\tilde{\bm{q}} = W \widehat{\bm{q}} \in \mathbb{R}^{L_{\rm k}},
	\end{equation}
	where $W\in\mathbb{R}^{{L_{\rm k}}\times {L_{\rm q}}}$ is a learnable projection. 
	The query is then compared with each key via a bilinear dot product, yielding the raw similarity score for $U_i$ as
	\begin{equation}
		\hat{s}_i = \bm{k}_i^{\!\top}\tilde{\bm{q}} .
	\end{equation}
	We normalize all scores across UAVs to obtain the normalized selection probability, given by
	\begin{equation}
		p_i = \frac{\exp(\hat{s}_i)}{\sum_{j=0}^N\exp(\hat{s}_j)} ,
	\end{equation}
	where $p_i\in(0,1)$ and $\sum_{j=0}^N p_j = 1$.
	An S-UAV transmits only if its normalized selection weight exceeds the threshold $s_{\rm th}$. In practice, $s_{\rm th}$ can be selected on a validation set to optimize downstream task performance.
	The source selection for $U_i$ can be expressed as
	\begin{equation}
		\Lambda_i \;=\;
		\begin{cases}
			1 , & p_i \ge s_{\rm th},\\
			0 , & p_i < s_{\rm th},
		\end{cases}
		\qquad i=1,\ldots,N ,
	\end{equation}
	where $s_{\rm th}\in(0,1)$ controls the cooperation level for who2com.
	Let $\bm{\epsilon}_{\mathrm{s}}$ denote the parameters of the semantic controller. If $\Lambda_i=1$, $U_i$ uploads its semantic feature over the uplink as
	\begin{equation}
		\widehat{\bm{V}}_i \;=\; \bm{h}_i \bm{V}_i \;+\; \bm{n}_i ,
		\label{eq:uplink-v}
	\end{equation}
	If $\Lambda_i=0$, $U_i$ remains silent, and no feature is transmitted. The communication group is $\mathcal{G}_t= \{U_0\}\cup\{\, U_i \mid \Lambda_i=1 \,\}$.

	\subsubsection{Data Fusion and Decoding}
	
	Only the T-UAV and the selected S-UAVs participate in fusion. Define the feature used for fusion as
	\begin{equation}
		\bm{V}_i^{\mathrm{fuse}} \;=\;
		\begin{cases}
			\bm{V}_0, & i=0,\\[2pt]
			\widehat{\bm{V}}_i, & i\neq 0 \text{ and } \ \Lambda_i=1,\\[2pt]
			\bm{0}, & i\neq 0 \text{ and } \ \Lambda_i=0,
		\end{cases}
	\end{equation}
	where the T-UAV’s own feature is locally obtained, each selected S-UAV contributes its channel-affected feature, and unselected S-UAVs contribute zeros. The fused representation can be expressed as
	\begin{equation}
		\widetilde{\bm{V}} = \sum_{i=0}^{N} \alpha_i \bm{V}_i^{\mathrm{fuse}},
	\end{equation}
	where the fusion weights $\alpha_i$ are defined as
	\begin{equation}
	\alpha_i \;=\;
	\begin{cases}
		[\hat{s}_0]_+, & i=0,\\
		[\hat{s}_i]_+\,\Lambda_i, & i=1,\ldots,N,
	\end{cases}
	\end{equation}
	where $[\hat{s}_i]_+=\max(0,\hat{s}_i)$ suppresses negative contributions while keeping unselected sources inactive via $\Lambda_i$.
	
	The semantic decoder produces the final output as
	\begin{equation}
		\bm{O}_0  =  f_{\rm{de}}(\widetilde{\bm{V}}, \bm{\epsilon}_{\mathrm{de}})\in\mathbb{R}^{n_{\rm{class}}\times H\times W},
	\end{equation}
	where $f_{\mathrm{de}}(\cdot,\cdot)$ and $\bm{\epsilon}_{\mathrm{de}}$ denote the mapping function and parameters of the semantic decoder.
	
	When2com is implicitly determined by the semantic controller. The controller outputs the normalized selection weights $\{p_i\}_{i=0}^{N}$ and applies who2com gating to obtain $\{\Lambda_i\}_{i=1}^{N}$. The decision of when2com is not optimized by an explicit auxiliary loss. Instead, it is implicitly learned from task supervision. When $\sum_{i=1}^{N}\Lambda_i = 0$, no S-UAV is activated. In this case, $\widetilde{\bm{V}}=\bm{V}_0$ and the T-UAV performs self-decoding as
	\begin{equation}
		\bm{O}_0 \;=\; f_{\mathrm{de}}(\bm{V}_0, \bm{\epsilon}_{\mathrm{de}})\in\mathbb{R}^{n_{\rm{class}}\times H\times W}.
		\label{decoder}
	\end{equation}
	Otherwise, the selected S-UAVs upload their features and the T-UAV performs cooperative fusion and decoding.
	
	\subsection{Resource Model}
	
	The proposed MC-DSA executes the task cooperatively. It forms a communication group via semantic matching and source selection, then exchanges only information expected to improve performance. We develop a communication resource model for bandwidth, latency, and energy that applies only when a communication group exists. 
    In practice, uplink transmissions follow TDMA. Each selected S-UAV sends its features to the T-UAV in a dedicated time slot, which prevents intra-group interference. As a result, the collaboration cost increases in direct proportion to the number of selected S-UAVs. Frames without a communication group incur no uplink cost, so the accounting naturally follows the semantic controller’s selection. 
	The query is a low-dimensional control message that carries only task intent for matching, whereas each selected S-UAV uploads a high-dimensional semantic feature tensor whose payload dominates the uplink cost, so we omit the query airtime.
	
	Each transmitting S-UAV uploads its semantic feature $\bm{V}_i$ with a fixed numeric precision. The per–S-UAV payload per frame is
	\begin{equation}
		\text{bits}_{\mathrm{sf}} \;=\; C_{\mathrm{feat}}\times (H/32)\times(W/32) \cdot \text{bpe},
	\end{equation}
	where $\text{bpe}$ is the number of bits per tensor element determined by the quantization format. Let $k=\sum_{i=1}^{N}\Lambda_i$ be the number of selected S-UAVs. The frame-level uplink payload can be expressed as
	\begin{equation}
		\text{bits}_{\mathrm{frame}} \;=\; \sum_{i=1}^{N}\Lambda_i\,\text{bits}_{\mathrm{sf}} \;=\; k\,\text{bits}_{\mathrm{sf}}.
	\end{equation}
	With a closed-loop update frequency $f$ in hertz, the aggregate bitrate requirement is
	\begin{equation}
		R_{\mathrm{req}} \;=\; f \cdot \text{bits}_{\mathrm{frame}} \quad [\mathrm{bit/s}] .
	\end{equation}
	Given the effective channel Signal-to-Noise Ratio (SNR) in linear scale denoted by $\mathrm{SNR}_{\rm{ch}}$, the Shannon spectral efficiency can be expressed as
	\begin{equation}
		\eta \;=\; \log_2\!\big(1+\mathrm{SNR}_{\rm{ch}}\big) \quad [\mathrm{bit/s/Hz}] ,
	\end{equation}
	Thus, the minimum bandwidth required to achieve $R_{\mathrm{req}}$ is
	\begin{equation}
		B_{\min} \;=\; \frac{R_{\mathrm{req}}}{\eta} \quad [\mathrm{Hz}].
	\end{equation}
	When no S-UAV is selected, $k=0$, one has $\text{bits}_{\mathrm{frame}}=0$ and $R_{\mathrm{req}}=0$.
	
	Per-frame latency is the sum of computation and communication terms, given by
	\begin{equation}
		T_{\mathrm{frame}} \;=\; T_{\mathrm{comp}} \;+\; T_{\mathrm{tx}} .
	\end{equation}
	The computation latency comprises three parts, semantic encoding, policy computation, and semantic fusion and decoding.
	\begin{equation}
		T_{\mathrm{comp}} = T_{\mathrm{enc}} + T_{\mathrm{policy}} + T_{\mathrm{dec}}.
	\end{equation}
	where $T_{\mathrm{enc}}$ is the per-frame time for all UAVs to perform semantic encoding. $T_{\mathrm{policy}}$ is the time to compute the when2com and who2com decisions, including query generation, key generation, and semantic similarity evaluation. $T_{\mathrm{dec}}$ is the time for semantic fusion and decoding at the T-UAV. Under TDMA, each selected S-UAV occupies the full bandwidth $B$ in its slot, so the communication latency is
	\begin{equation}
		T_{\mathrm{tx}} \;=\; \sum_{i=1}^{N}\Lambda_i\,\frac{\text{bits}_{\mathrm{sf}}}{\mathcal{C}}
		\;=\; \frac{\text{bits}_{\mathrm{frame}}}{\mathcal{C}}, \quad
		\mathcal{C}=B\,\eta,
	\end{equation}
	and the S-UAV transmission time is $T_{\mathrm{tx,sf}}=\text{bits}_{\mathrm{sf}}/\mathcal{C}$, hence $T_{\mathrm{tx}}=k\,T_{\mathrm{tx,sf}}$. When $k=0$, $T_{\mathrm{tx}}=0$ and $T_{\mathrm{frame}}=T_{\mathrm{comp}}$.
	
	The per-frame energy consists of computation energy and communication energy, given by
	\begin{equation}
		E_{\mathrm{frame}} \;=\; E_{\mathrm{comp}} \;+\; E_{\mathrm{comm}} .
	\end{equation}
	The computation energy is given by
	\begin{equation}
		E_{\mathrm{comp}}
		= P_{\mathrm{comp}}^{T}\,T_{\mathrm{comp}}^{T}
		+ \sum_{i=1}^{N} P_{\mathrm{comp},i}^{S}\,T_{\mathrm{comp},i}^{S}.
		\label{eq:ecomp}
	\end{equation}
	Under TDMA uplink, only selected S-UAVs transmit and the T-UAV receives. The total receive time at the T-UAV is \(T_{\mathrm{rx}}=\sum_{i=1}^{N} T_{\mathrm{tx},i}\). The communication energy can be expressed as
	\begin{equation}
		E_{\mathrm{comm}}
		= P_{\mathrm{rx}}^{T}\,T_{\mathrm{rx}}
		+ \sum_{i=1}^{N} \Lambda_i\, P_{\mathrm{tx},i}^{S}\, T_{\mathrm{tx},i}.
		\label{eq:ecomm}
	\end{equation}
	When no S-UAV is selected, $T_{\mathrm{rx}}=0$ and $E_{\mathrm{comm}}=0$.
	
	\subsection{Loss Function}
	
	The proposed MC-DSA employs the cross-entropy (CE) loss between the predicted class probabilities and the ground-truth labels. This differentiable loss provides stable supervision under stochastic channels, directly optimizes segmentation accuracy, and is well suited to end-to-end learning.
	
	For $\bm{O}_0$, let $p_j(c)$ denote the softmax probability of class $c$ at pixel $j$.
	Ground-truth labels are $\bm{L}_0\in\mathbb{R}^{H\times W}$. The training objective is the average CE
	\begin{equation}
		L_{\mathrm{CE}}
		= -\frac{1}{|\mathcal{J}|}\sum_{j\in\mathcal{J}}
		\sum_{c=1}^{n_{\rm{class}}}\mathbf{1}\!\big(\bm{L}_{0,j}=c\big)\,\log p_j(c).
		\label{eq:loss_ce}
	\end{equation}
	where $\mathbf{1}(\cdot)$ is the indicator function.	
	For model selection, we monitor the validation task accuracy $A_{\mathrm{val}}^{(t)}$ at epoch $t$. With patience $P$ and a staged learning-rate schedule with $S$ stages, define the best accuracy within the most recent $P$ epochs as
	\begin{equation}
		B_P(t)=\max_{t'\in[t-P+1,\,t]} A_{\mathrm{val}}^{(t')}.
	\end{equation}
	If $B_P(t)\le A_{\mathrm{best}}$ and the current stage index $s<S-1$, we reduce the learning rate and reset patience. If $B_P(t)\le A_{\mathrm{best}}$ and $s=S-1$, we stop training early. Otherwise, when $A_{\mathrm{val}}^{(t)}>A_{\mathrm{best}}$ we update $A_{\mathrm{val}}^{(t)}$ to $A_{\mathrm{best}}$ and reset patience. $A_{\mathrm{best}}$ is the best validation accuracy observed so far and $s\in\{0,\dots,S-1\}$ is the current learning-rate stage index. This policy aligns checkpoint selection with the deployment objective, since the downstream user cares about segmentation reliability under the given bandwidth and SNR.
	
	From a computational perspective, the proposed MC-DSA framework is scalable with respect to the number of UAVs. Each UAV independently performs semantic encoding, resulting in a computational cost of $O(N \cdot C_{\text{enc}})$, where $C_{\text{enc}}$ denotes the encoding complexity of a single UAV. The semantic matching process operates on fixed-dimensional query–key representations, leading to a complexity of $O(N \cdot L_{\rm q} \cdot L_{\rm k})$. 
	This term mainly comes from query/key generation and semantic similarity evaluation in the policy branch, and grows with the number of S-UAVs while remaining moderate due to the compact representations.
	Moreover, cooperative transmission and feature fusion are conducted only among a subset of selected UAVs, which results in a fusion complexity of $O(k \cdot C_{\text{fuse}} + C_{\text{dec}})$. Consequently, the overall computational complexity of the proposed MC-DSA can be expressed as $O(N \cdot C_{\text{enc}} + N \cdot L_{\rm q} \cdot L_{\rm k} + k \cdot C_{\text{fuse}} + C_{\text{dec}})$.
	
	\section{Algorithm Design}
	
	In this section, we propose algorithms for centralized training and testing-time inference. The training algorithm learns a semantic matching policy together with task completion. The testing algorithm evaluates the learned policy under TDMA uplink over a sweep of SNRs.
	
	\vspace{-1 mm}
	\subsection{Training Algorithm}
	\vspace{-1 mm}

	\begin{algorithm}[b!]
		\caption{\textbf{Training Algorithm}}
		\label{training}
		\begin{algorithmic}[1]
			\small
			\State \textbf{Input:} training set $\mathcal{D}_{\mathrm{tr}}=\{(\bm{i}_0,\bm{i}_1,\dots,\bm{i}_N,\bm{L}_0)\}$.
			\State \textbf{Output:} training parameters $\bm{\epsilon}_{\mathrm{en}},
			\bm{\epsilon}_{\mathrm{pen}},
			\bm{\epsilon}_{\mathrm{q}},
			\bm{\epsilon}_{\mathrm{k}},
			\bm{\epsilon}_{\mathrm{s}}, \bm{\epsilon}_{\mathrm{de}}$.
			
			\State Initialize all parameters,
			\For{not converged}
			\Statex \qquad  \textit{// Semantic encoder}
			
			\State 
			$\bm{V}_i \gets f_{\mathrm{en}}(\bm{i}_i,\bm{\epsilon}_{\mathrm{en}})$ for $i = 0, \ldots, N$,
			
			\Statex \qquad \textit{// Q/K generator}
			
			\State
			$\bm{V}_{\rm qk}^i \gets f_{\mathrm{pen}}(\bm{i}_i,\bm{\epsilon}_{\mathrm{pen}})$ for $i = 0, \ldots, N$,
			\State $\bm{q}\gets f_{\rm q}(\bm{V}_{\rm qk}^0,\bm{\epsilon}_{\rm q})$,
			
			\State $\bm{k}_i\gets f_{\rm k}(\bm{V}_{\rm qk}^i,\bm{\epsilon}_{\rm k})$ for $i = 0, \ldots, N$,
			
			\Statex \qquad \textit{// Channel }
			\State $\widehat{\bm{q}} \gets \mathrm{Channel}(\bm{q};\mathrm{SNR}_{\rm ch})$,
			
			\Statex \qquad \textit{// Semantic controller}
			\State $\hat{s}_i \gets f_{\rm s}(\widehat{\bm{q}} ,\bm{k}_i;\bm{\epsilon}_{\rm s})$ for $i = 0, \ldots, N$,
			
			\State  $\alpha_i^{\rm soft} \gets [\hat{s}_i]_+$ for $i = 0, \ldots, N$,
			
			\Statex \qquad \textit{// Channel }
			\State $\widehat{\bm{V}}_i \gets \mathrm{Channel}(\bm{V}_i;\mathrm{SNR}_{\rm ch})$ for $i=1, \ldots, N$,
			
			\Statex \qquad \textit{// Semantic decoder}
			\State $\bm{V}_0 \gets \bm{V}_0$, $\bm{V}_i^{\rm fuse} \gets \widehat{\bm{V}}_i$,
			\State $\widetilde{\bm{V}} \gets \sum_{i=0}^{N} \alpha_i^{\rm soft}\,\bm{V}_i^{\rm fuse}$,
			\State $\bm{O}_0 \gets f_{\mathrm{de}}(\widetilde{\bm{V}},\bm{\epsilon}_{\mathrm{de}})$,
			
			\Statex \qquad \textit{// Loss and update}
			\State Compute loss $\mathcal{L}\gets \mathcal{L}_{\mathrm{CE}}(\bm{O}_0,\bm{L}_0)$,
			\State Update training parameters by gradient descent,
			
			\Statex \qquad \textit{// Validation monitoring and early stopping}
			\State Compute $B_P(t)=\max_{t'\in[t-P+1,\,t]}A_{\mathrm{val}}^{(t')}$ ,
			\If{$A_{\mathrm{val}}^{(t)} > A_{\mathrm{best}}$}
			\State $A_{\mathrm{best}} = A_{\mathrm{val}}^{(t)}$, save checkpoint,
			\EndIf
			\If{$B_P(t)\le A_{\mathrm{best}}$}
			\If{$s < S-1$}
			\State Reduce learning rate, $s \gets s+1$, \textbf{continue}
			\Else
			\State \textbf{break}
			\EndIf
			\EndIf
			
			\EndFor
			\State Save the trained parameters.
		\end{algorithmic}
	\end{algorithm}

	The training algorithm aims to optimize the parameters of key components. The training setup assumes access to synchronized multi-view data and labels. The input is the training set $\mathcal{D}_{\mathrm{tr}}=\{\bm{i}_0,\bm{i}_1,\ldots,\bm{i}_N,\bm{L}_0\}$. The trainable parameters are $\bm{\epsilon}_{\mathrm{en}},
	\bm{\epsilon}_{\mathrm{pen}},
	\bm{\epsilon}_{\mathrm{q}},
	\bm{\epsilon}_{\mathrm{k}},
	\bm{\epsilon}_{\mathrm{s}}, 
	\bm{\epsilon}_{\mathrm{de}}$. All components are optimized jointly by backpropagation.
	
	The shared semantic encoder produces  $\bm{V}_i$ for $i=0,\ldots,N$. Policy encoders yield $\bm{V}_{\rm qk}^i$ for $i=0,\ldots,N$. The T-UAV maps $\bm{V}_{\rm qk}^0$ to a query $\bm{q}$ and each UAV maps $\bm{V}_{\rm qk}^i$ to a key $\bm{k}_i$ for $i=0,\ldots,N$. The query is broadcast over the wireless channel, yielding the received signal $\widehat{\bm{q}}$. The semantic controller computes similarity scores, which are normalized to probabilities. Nonnegative soft weights are set as $\alpha_i^{\rm soft}$. Each S-UAV feature is transmitted over the wireless channel to obtain $\widehat{\bm{V}}_i$, from which the fused feature $\bm{V}_i^{\rm fuse}$ is formed.
	The fused representation is formed $\widetilde{\bm{V}}$ as a soft weighted sum, and the semantic decoder produces the output $\bm{O}_0$.
	
	The loss is the $L_{\mathrm{CE}}$ between the semantic decoder output and the ground-truth label. Gradients update the parameters $\bm{\epsilon}_{\mathrm{en}},
	\bm{\epsilon}_{\mathrm{pen}},
	\bm{\epsilon}_{\mathrm{q}},
	\bm{\epsilon}_{\mathrm{k}},
	\bm{\epsilon}_{\mathrm{s}}, 
	\bm{\epsilon}_{\mathrm{de}}$. Through the loss function, the algorithm can guide the parameter optimization of these components, thereby improving the performance of the entire system in the image segmentation task. 
	After each epoch a validation metric is computed on a held-out set, a patience mechanism tracks the best checkpoint, and early stopping is triggered when the metric stops improving. Finally, the algorithm saves the trained network parameters.
	
	\begin{algorithm}[b!]
		\caption{\textbf{Testing Algorithm}}
		\label{Testing}
		\begin{algorithmic}[1]
			\small
			\State \textbf{Input:} test set $\mathcal{D}_{\mathrm{te}}$, trained parameters $\bm{\epsilon}_{\mathrm{en}},
			\bm{\epsilon}_{\mathrm{pen}},
			\bm{\epsilon}_{\mathrm{q}},
			\bm{\epsilon}_{\mathrm{k}},
			\bm{\epsilon}_{\mathrm{s}}, \bm{\epsilon}_{\mathrm{de}}$; 
			selection threshold $s_{\rm th}$; 
			SNR set $\mathcal{S}$
			
			\State \textbf{Output:} per-SNR task metrics
			
			\For{each SNR $s \in \mathcal{S}$}
			\State reset metric accumulators,
			
			\For{each frame}
			
			\Statex \qquad \textit{// T-UAV}
			\State $\bm{V}_0 \gets f_{\mathrm{en}}(\bm{i}_0,\bm{\epsilon}_{\mathrm{en}})$,
			\State $\bm{V}_{\rm qk}^0 \gets f_{\mathrm{pen}}(\bm{i}_0,\bm{\epsilon}_{\mathrm{pen}})$,
			\State 
			$\bm{q}\gets f_{\rm q}(\bm{V}_{\rm qk}^0,\bm{\epsilon}_{\rm q})$,
			\State $\widehat{\bm{q}} \gets \mathrm{Channel}(\bm{q};\mathrm{SNR}_{\rm ch})$,
			\State 
			$\bm{k}_0 \gets f_{\rm k}(\bm{V}_{\rm qk}^0,\bm{\epsilon}_{\rm k})$,
			\State
			$s_0 \gets f_{\rm s}(\bm{q},\bm{k}_0;\bm{\epsilon}_{\rm s})$,

			\Statex \qquad \textit{// S-UAVs}
			
			\For{$i=1$ \textbf{to} $N$ \textbf{in parallel}}
			\State $\bm{V}_i \gets f_{\mathrm{en}}(\bm{i}_i,\bm{\epsilon}_{\mathrm{en}})$,
			\State $\bm{V}_{\rm qk}^i \gets f_{\mathrm{pen}}(\bm{i}_i,\bm{\epsilon}_{\mathrm{pen}})$,
			\State 
			$\bm{k}_i \gets f_{\rm k}(\bm{V}_{\rm qk}^i,\bm{\epsilon}_{\rm k})$,
			\State $p_i \gets \hat{s}_i \gets f_{\rm s}(\widehat{\bm{q}},\bm{k}_i;\bm{\epsilon}_{\rm s})$,
			\State 
			$\Lambda_i \gets \mathbb{I}\{ p_i \ge s_{\rm th} \}$,
			\If{$\Lambda_i=1$} 
			\State $\widehat{\bm{V}}_i \gets \mathrm{Channel}\big(\bm{V}_i;\ \mathrm{SNR}_{\rm ch}=s\big)$,
			\EndIf
			\EndFor
			\Statex \qquad \textit{//  T-UAV}
			\If{$\sum_{i=1}^{N}\Lambda_i=0$}
			\State $\widehat{\bm{O}}_0 \gets f_{\mathrm{de}}(\bm{V}_0,\bm{\epsilon}_{\mathrm{de}})$, update metrics,
			\State \textbf{continue} to next frame,
			\Else
			\State $\bm{V}_0^{\mathrm{fuse}} \gets \bm{V}_0$,
			\If{$\Lambda_i=1$}
			\State $\bm{V}_i^{\mathrm{fuse}} \gets \widehat{\bm{V}}_i$,
			\EndIf
			\State $\alpha_0 \gets [s_0]_+$,\quad $\alpha_i \gets [\hat{s}_i]_+\,\Lambda_i$ for $i=1,\dots,N$,
			\State $\widetilde{\bm{V}} \gets \sum_{i=0}^{N} \alpha_i \bm{V}_i^{\mathrm{fuse}}$,
			\State $\widehat{\bm{O}}_0 \gets f_{\mathrm{de}}(\widetilde{\bm{V}},\bm{\epsilon}_{\mathrm{de}})$,
			\State update metrics,
			\EndIf
			\EndFor
			\EndFor
			\State \Return per-SNR metrics.
		\end{algorithmic}
	\end{algorithm}
	
	\subsection{Testing Algorithm}
	
	In the testing phase, we assess the proposed MC-DSA across different SNRs. The when2com decision is implicit and realized by thresholding the learned scores, so only selected S-UAVs participate in transmission and fusion. The inputs are the test set $\mathcal{D}_{\mathrm{te}}$, the trained parameters $\bm{\epsilon}_{\mathrm{en}},
	\bm{\epsilon}_{\mathrm{pen}},
	\bm{\epsilon}_{\mathrm{q}},
	\bm{\epsilon}_{\mathrm{k}},
	\bm{\epsilon}_{\mathrm{s}}, \bm{\epsilon}_{\mathrm{de}}$, the threshold $s_{\rm th}$, and an SNR set $\mathcal{S}$. The output is the metrics of the image semantic segmentation.

	For each SNR $s\in\mathcal{S}$ and for each frame, the T-UAV encodes $\bm{i}_0$ to $\bm{V}_0$, forms a query $\bm{q}$, and broadcasts it through the channel to obtain $\widehat{\bm{q}}$. Each S-UAV encodes its view $\bm{i}_i$ to $\bm{V}_i$. Each UAV maps the image to a key $\bm{k}_i$, computes a score $\hat{s}_i$, and applies hard selection $\Lambda_i$. If no S-UAV is selected, the T-UAV performs self-only decoding. Otherwise each selected S-UAV uploads $\bm{V}_i$. At T-UAV, fusion uses $\alpha_i$ with $\bm{V}_i^{\rm fuse}$. The semantic decoder produces the prediction $\widehat{\bm{O}}_0$, and task metrics are accumulated and then summarized for each SNR.
	
	\section{Simulation Experiment}
	
	In this section, we introduce the simulation setup, including parameter settings, performance metrics, and benchmarks, analyze the simulation results, and validate the feasibility of the proposed MC-DSA.
	
	\subsection{Simulation Setting}
	
	Tab. \ref{tabsettings} summarizes the parameter settings. Experiments are conducted on the AirSim–SRMS dataset with inputs resized to $512{\times}512$ \cite{liu2020when2com,airsim}. The $L_{\rm q}$ is 8 and the $L_{\rm k}$ is 1024. The training objective is pixel-wise CE. Communication is emulated using Additive White Gaussian Noise (AWGN), Rician fading, and Rayleigh fading channel models with SNRs \{-20,-10,0,10,20\}\,dB. Unless otherwise stated, the communication bandwidth is set to 10 MHz \cite{mUAV1} and the frame rate to 10 Hz. The power budget assumes on-board compute power 15 W and radio transmit or receive powers 10 W each \cite{guan2021energy,li2020energy}. The implementation is in Python using PyTorch with CUDA data parallelism when available, and TensorBoardX is used for logging. We repeat the training multiple times and demonstrate the mean performance over the repeated runs.
	Optimization uses Adam with an initial learning rate of $1{\times}10^{-4}$ and a staged decay to $5{\times}10^{-6}$. We employ a mini-batch of 6 with 4-step gradient accumulation, effective batch size 24, train for up to 120 epochs, and validate after every epoch. Early stopping monitors validation Overall Accuracy (OAcc) with a patience 6 and the best checkpoint is retained.

	\begin{table}[b!]
		\centering
		\caption{Parameter settings for the proposed MC-DSA.}
		\label{tabsettings}
		\begin{tabular}{ll}
			\hline
			\textbf{Item} & \textbf{Setting} \\
			\hline
			Platform & Python \\
			Dataset  & AirSim-SRMS \\
			Image resolution $H{\times}W$ & $512{\times}512$ \\
			UAV number $N$ & $5$\\
			Query length $L_{\rm q}$ & $8$ \\
			Key length $L_{\rm k}$ & $1024$ \\
			Loss & CE \\
			Optimizer & Adam \\
			LR schedule & $1{\times}10^{-4}\!\rightarrow\!5{\times}10^{-6}$  \\
			Effective batch &  $24$ \\
			Max epochs & $120$ \\
			Early stopping patience & $6$ \\
			Channel models & AWGN, Rician fading, Rayleigh fading \\
			Test SNR (dB) & $\{-20,\ -10,\ 0,\ 10,\ 20\}$ \\
			Bandwidth & $10$\,MHz \\
			Frame rate & $10$\,Hz \\
			Compute power & $15$\,W \\
			Tx/Rx power & $10$\,W  \\
			\hline
		\end{tabular}
	\end{table}

	To comprehensively evaluate the proposed MC-DSA, we report two classes of metrics: perception quality and system efficiency. The former measures the semantic segmentation performance, while the latter quantifies the communication and computation overhead.
	For perception quality, we report OAcc, mean Accuracy (mAcc), Frequency-Weighted Intersection-over-Union (FWIoU), and mean IoU (mIoU) \cite{mIOU}. OAcc characterizes correctness over all valid pixels. mAcc is the unweighted average of per-class pixel accuracies, mitigating bias from class imbalance. FWIoU is the pixel-frequency-weighted average of per-class IoUs, reflecting overall overlap quality under the ground-truth pixel distribution. mIoU is the unweighted average of per-class IoUs and measures the overlap between predictions and annotations. 
	For system efficiency, we use bandwidth requirement, cumulative latency, and cumulative energy. Bandwidth requirement is the link bandwidth needed to sustain the target throughput when transmitting features from selected S-UAVs. Cumulative latency is the accumulated latency from the first frame to the current frame. Cumulative energy is the accumulated energy from the first frame to the current frame.
	
	We evaluate the proposed MC-DSA with a set of benchmarks under a unified protocol. All methods use the same training data, loss, optimizer, policy structure, fusion rule, and semantic decoder. The policy branch mirrors the semantic encoder backbone and is trained jointly with the task network for every method. Performance differences therefore arise only from the encoder choice or the cooperation mode.
	
	\begin{enumerate}
		
		\item MC-DSA@Cooperative: when the T-UAV needs support, it broadcasts a query. Each S-UAV evaluates relevance, and only the selected ones transmit semantic features. The T-UAV fuses its local feature with the received features and then decodes to produce the output.
		
		\item MC-DSA@Standalone: when the T-UAV observation is sufficient, no cooperation occurs. All $\Lambda_i$ are zero and there is no uplink transmission. The semantic decoder operates only on the local feature $\bm{V}_0$ to report the single-view inference performance.
		
		\item ResNet \cite{ResNet}: the semantic encoder is a standard ResNet backbone that produces a fixed spatial feature grid for the semantic decoder. This benchmark represents a widely used convolutional encoder without attention.
		
		\item ECANet \cite{ECA–Net}: the semantic encoder is a ResNet augmented with efficient channel attention applied within residual blocks. The output spatial grid matches the common setting and feeds the same semantic decoder.
		
		\item SegNet \cite{SegNet}: the semantic encoder follows an encoder pyramid built from Conv blocks with progressive downsampling to the fixed feature grid. This benchmark represents a pure convolutional alternative without residual connections or self-attention.

		\item T-UAV view: only the T-UAV observation is used for task execution, without any S-UAV cooperation.
		
		\item Random select: the T-UAV randomly selects one S-UAV for cooperation. The selected S-UAV transmits its semantic feature to the T-UAV.
		
		\item Random fusion: all available S-UAV features are transmitted, and the T-UAV fuses them with randomly assigned fusion weights before decoding.

	\end{enumerate}

	\subsection{Simulation Results}
	
	\subsubsection{Analysis of Cooperative Communication}
	
	\begin{figure}[b!]
		\centering
		
		\subfloat[]{\includegraphics[width=0.99\linewidth]{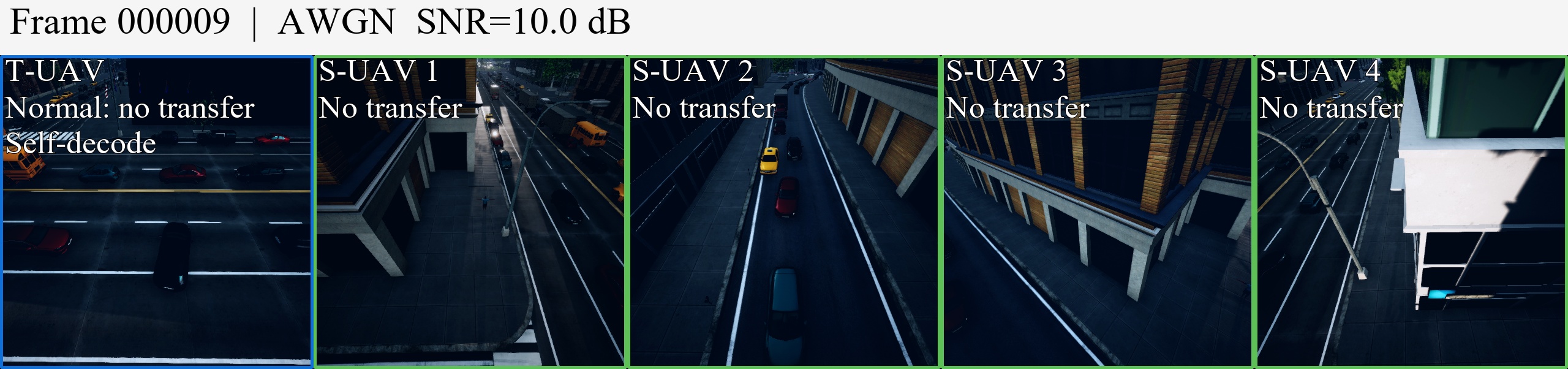}\label{fig:vis:a}}\\[-2pt]
		\subfloat[]{\includegraphics[width=0.99\linewidth]{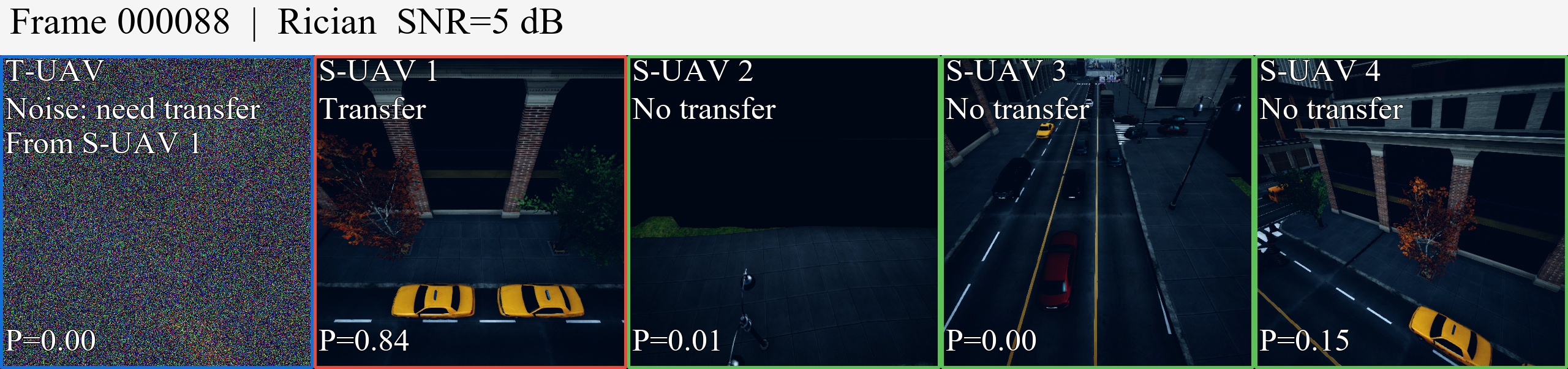}\label{fig:vis:b}}\\[-2pt]
		\subfloat[]{\includegraphics[width=0.99\linewidth]{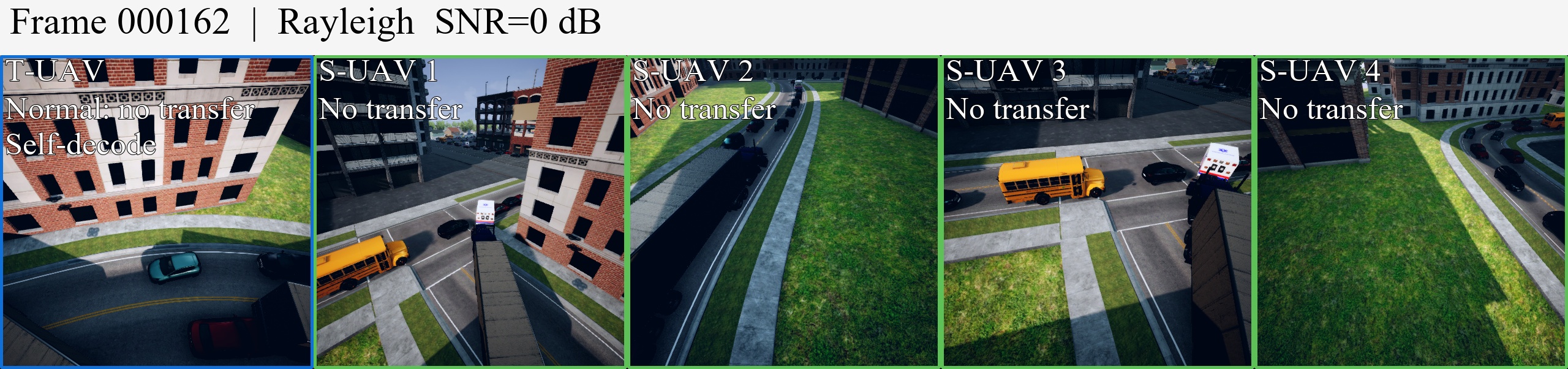}\label{fig:vis:c}}\\[-2pt]
		\subfloat[]{\includegraphics[width=0.99\linewidth]{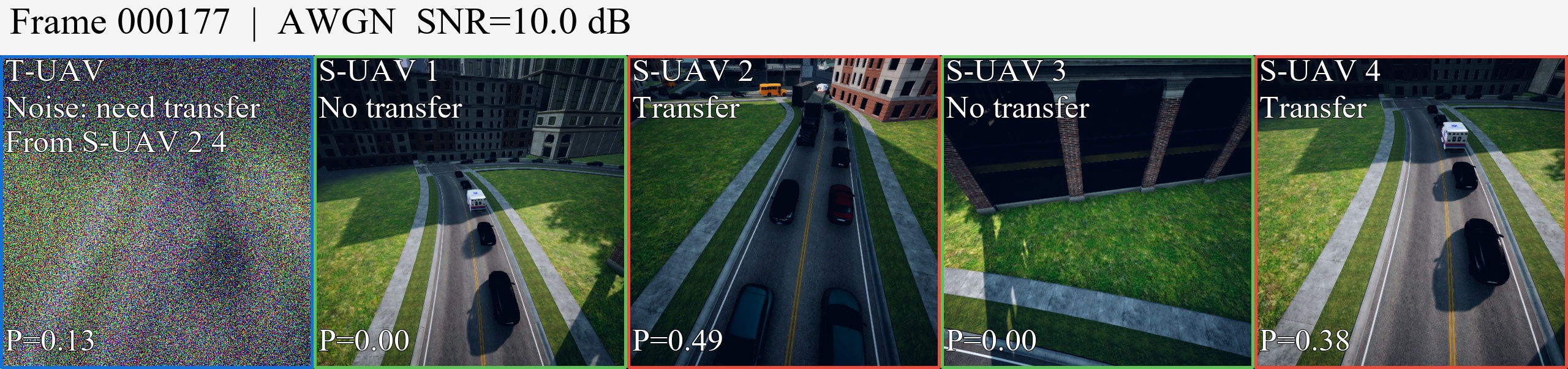}\label{fig:vis:d}}\\[-2pt]
		\subfloat[]{\includegraphics[width=0.99\linewidth]{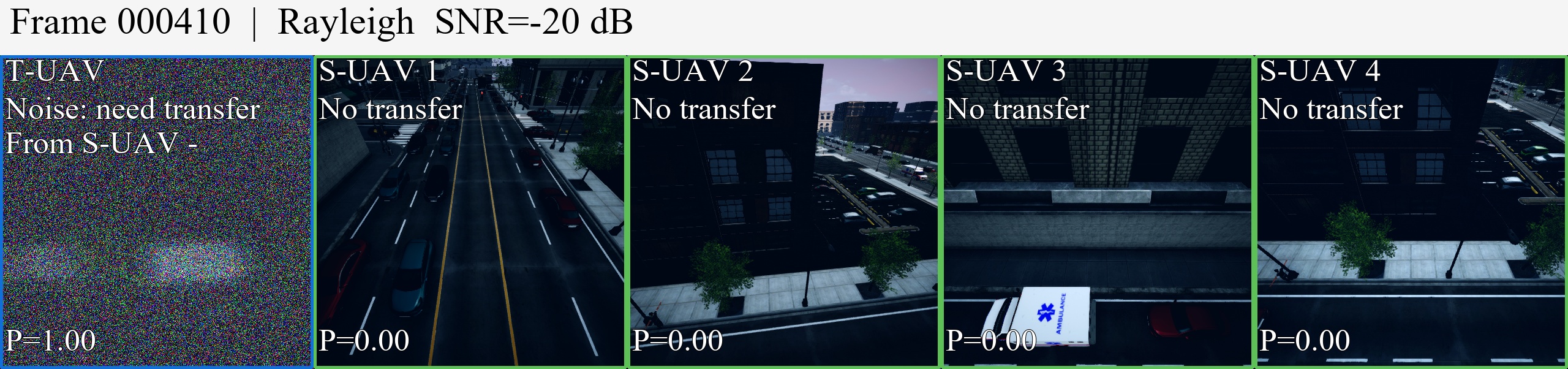}\label{fig:vis:e}}\\[-2pt]
		\subfloat[]{\includegraphics[width=0.99\linewidth]{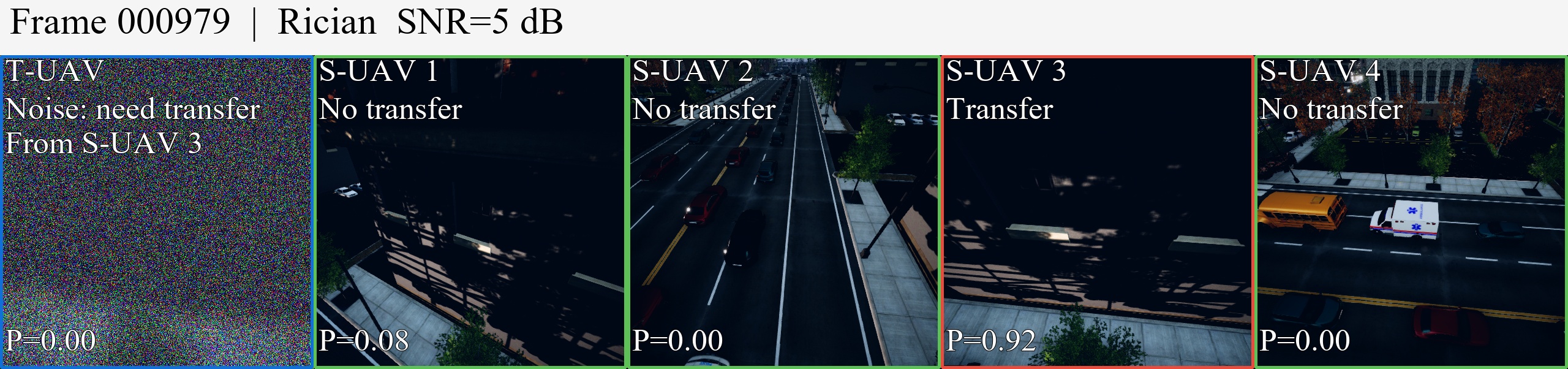}\label{fig:vis:f}}
		
		\caption{Visualization of task execution with the proposed MC-DSA.}
		\label{ResultVisualization}
	\end{figure}
	
	\begin{figure*}[t]
		\centering
		\includegraphics[width=0.99\textwidth]{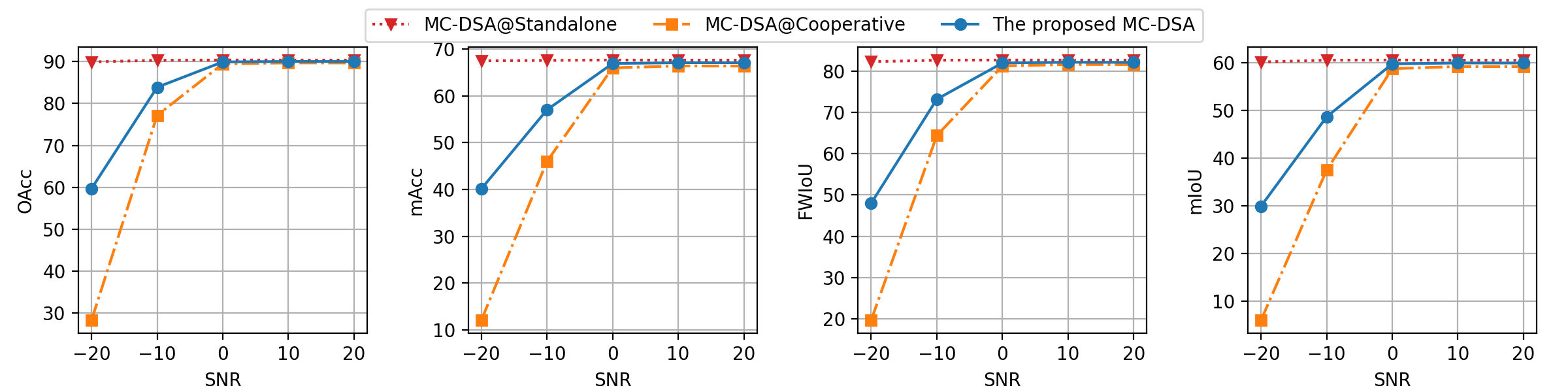}
		\caption{Segmentation performance of different MC-DSA operating modes under the AWGN channel.}
		\label{metric_all_mode}
	\end{figure*}
	
	\begin{figure*}[t]
		\centering
		\includegraphics[width=0.99\textwidth]{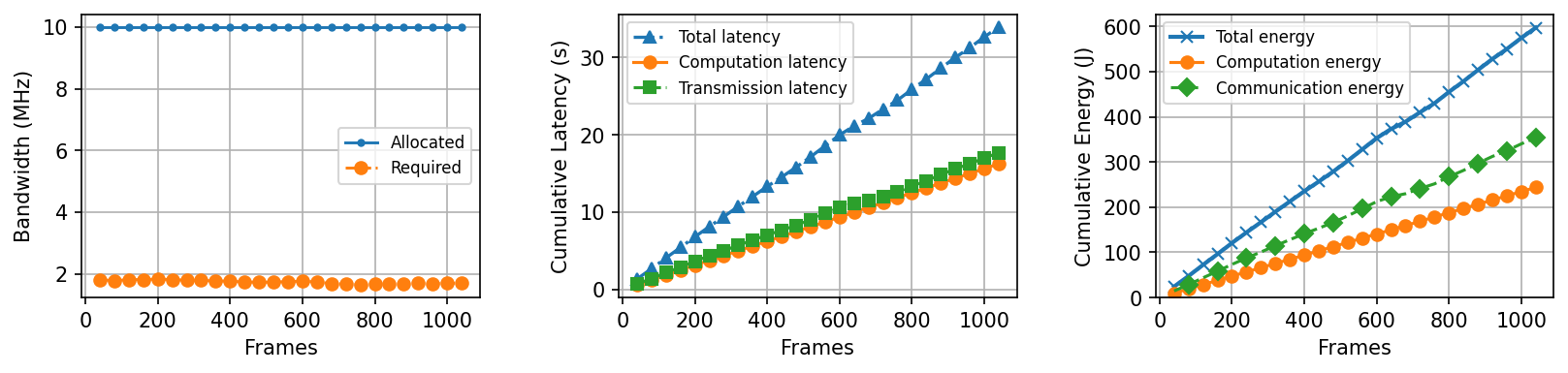}
		\caption{Resource performance of the proposed MC-DSA under the Rayleigh fading channel.}
		\label{overall_kpi_cum_bte}
	\end{figure*}

	\begin{figure*}[t]
		\centering
		\subfloat[AWGN\label{Bench_AWGN}]%
		{\includegraphics[width=0.99\linewidth]{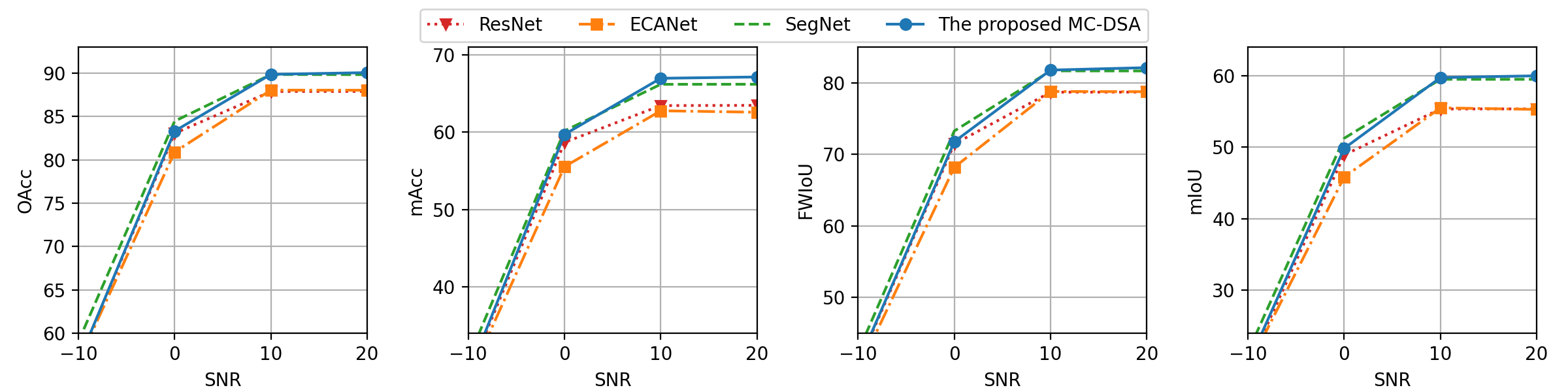}}\\[2pt]
		\subfloat[Rician fading\label{Bench_Rician}]%
		{\includegraphics[width=0.99\linewidth]{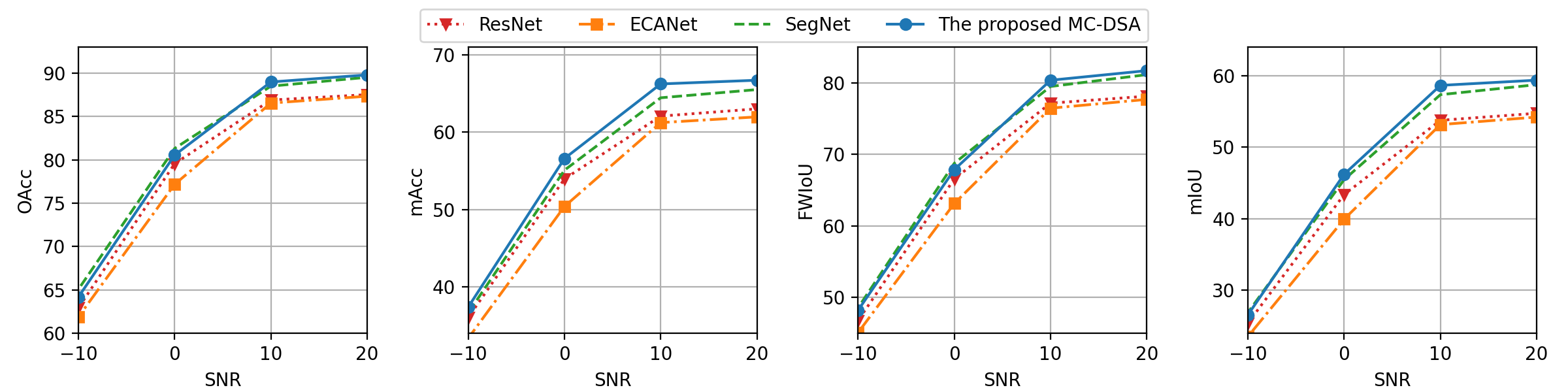}}\\[2pt]
		\subfloat[Rayleigh fading\label{Bench_Rayleigh}]%
		{\includegraphics[width=0.99\linewidth]{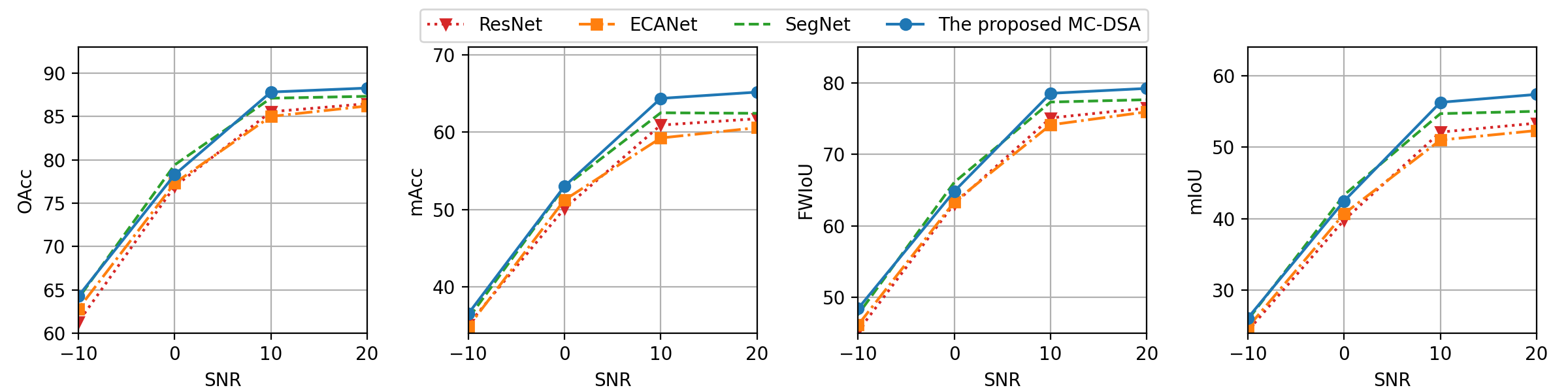}}
		\caption{Segmentation performance of different systems under different channels.}
		\label{channelsmetric}
	\end{figure*}
	
	\begin{figure}[t]
		\centering
		\subfloat[AWGN\label{selection_acc_AWGN_snr}]%
		{\includegraphics[width=0.99\linewidth]{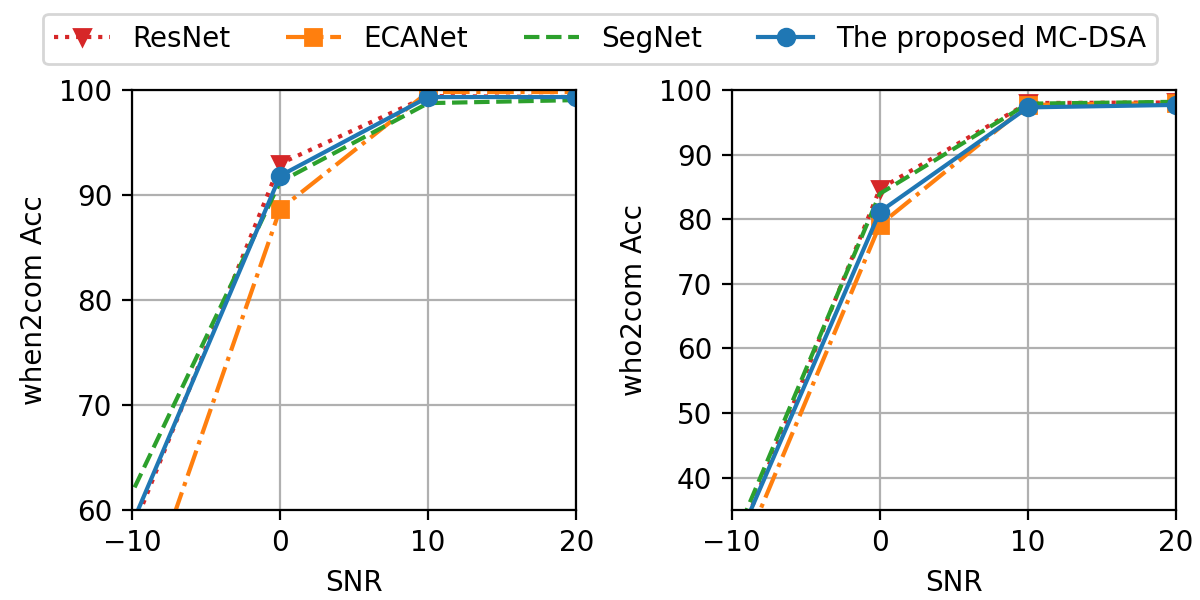}}\\[2pt]
		\subfloat[Rician fading\label{selection_acc_Rician_snr}]%
		{\includegraphics[width=0.99\linewidth]{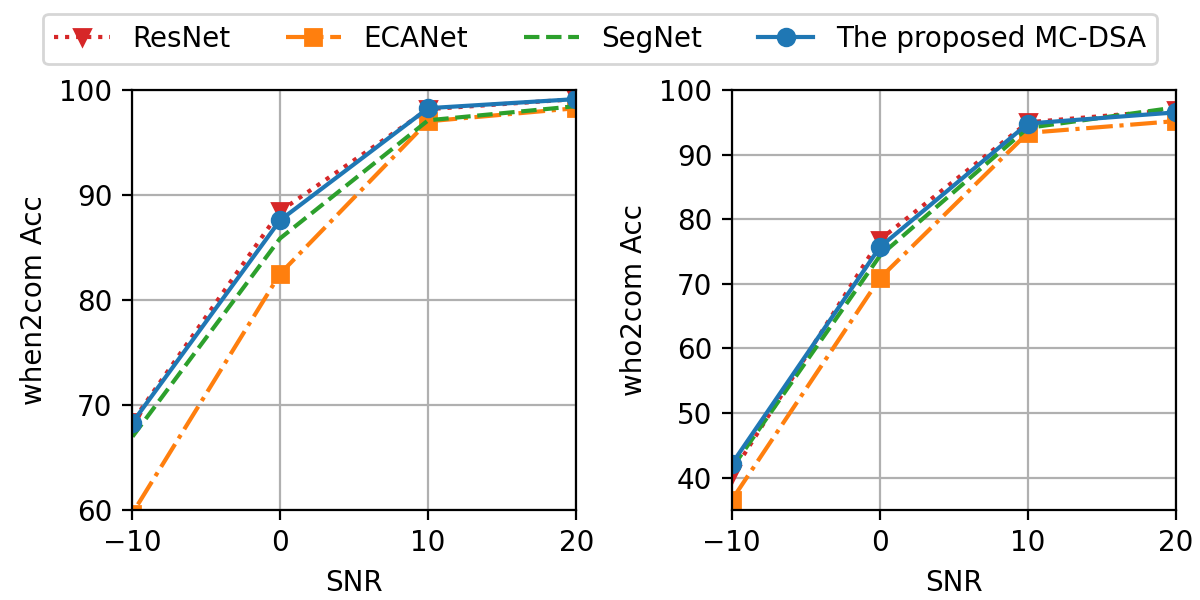}}\\[2pt]
		\subfloat[Rayleigh fading\label{selection_acc_Rayleigh_snr}]%
		{\includegraphics[width=0.99\linewidth]{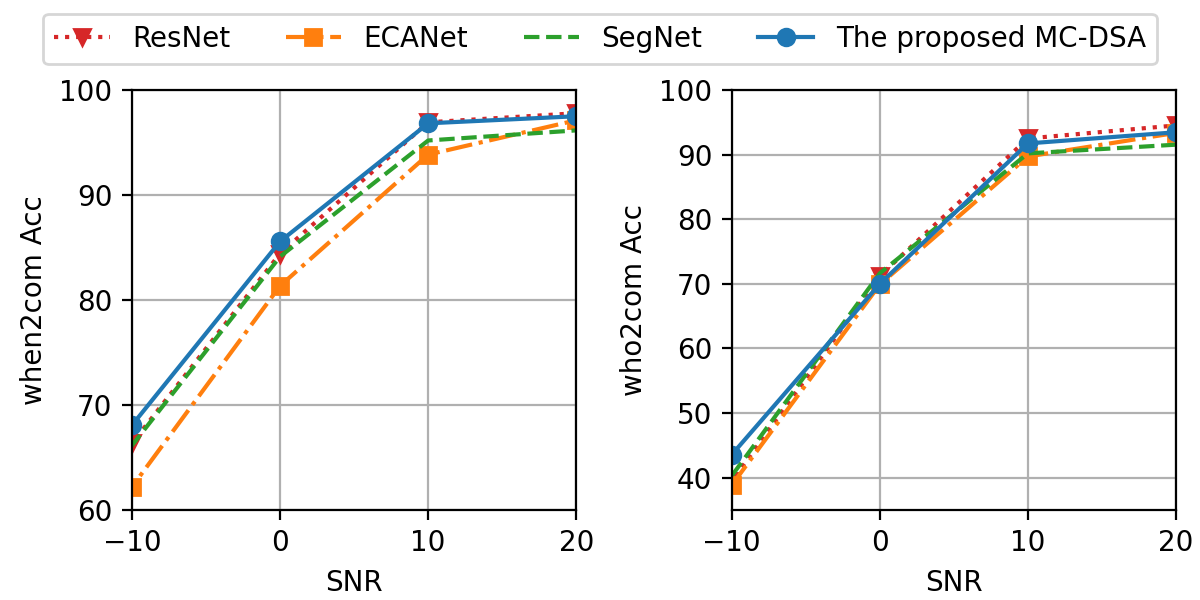}}
		\caption{Selection accuracy of different systems under different channels.}
		\label{channelsselacc}
	\end{figure}
	
	Fig. \ref{ResultVisualization} illustrates how the proposed MC-DSA switches between self-decoding and cooperation under different channels and SNRs. In each frame, the left column shows the T-UAV view and the remaining columns show the four S-UAVs. The lower-left value of each image denotes the normalized relevance probability used for transmission decision.
	In (a) AWGN at 10 dB and (c) Rayleigh fading at 0 dB, the T-UAV view is clear and sufficient, so it self-decodes and all S-UAVs remain silent. In (b) Rician fading at 5 dB, the T-UAV view is degraded and the policy activates cooperation by selecting only S-UAV 1 with probability 0.84, while the other S-UAVs are suppressed due to low relevance. In (d) AWGN at 10 dB, it selects S-UAV 2 with probability 0.49 and S-UAV 4 with probability 0.38. In (f) Rician fading at 5 dB, it selects S-UAV 3 with probability 0.92. In (e) Rayleigh fading at -20 dB, the T-UAV view is severely corrupted and all S-UAVs have near zero relevance, so no useful S-UAV is found. Overall, the visualization confirms that MC-DSA activates cooperation only when needed and with minimal transmissions, demonstrating adaptability and robustness across different channel conditions.

	Fig. \ref{metric_all_mode} compares the segmentation performance of the proposed MC-DSA under the AWGN channel across different SNRs, using OAcc, mAcc, FWIoU, and mIoU. In this experiment, the two benchmarks correspond to distinct operating modes of MC-DSA under the same settings.	
	At an SNR of -20 dB, MC-DSA@Cooperative achieves about 28\% OAcc, 12\% mAcc, 20\% FWIoU, and 6\% mIoU, whereas the proposed MC-DSA reaches about 60\% OAcc, 40\% mAcc, 48\% FWIoU, and 30\% mIoU. At -10 dB, the proposed MC-DSA still maintains clear gains over MC-DSA@Cooperative. At 0 dB, it reaches 90\% OAcc, 68\% mAcc, 82\% FWIoU, and 60\% mIoU, closely approaching MC-DSA@Standalone, which remains near 90\% OAcc, 69\% mAcc, 82\% FWIoU, and 61\% mIoU across the whole range.	
	Overall, the proposed MC-DSA consistently outperforms MC-DSA@Cooperative on all metrics from low to high SNR and rapidly approaches the MC-DSA@Standalone upper bound as the channel improves. This is because MC-DSA@Standalone relies only on the T-UAV observation and avoids transmission-induced distortion, while MC-DSA@Cooperative always requires multi-UAV feature transmission over the wireless channel, which may degrade the fused representation.	
	The superiority of the proposed MC-DSA lies in its adaptive semantic matching and source selection mechanism. When the T-UAV observation is sufficient, it operates in standalone mode and avoids unnecessary transmission. Otherwise, it collaborates with S-UAVs to exploit complementary observations. In this way, the policy branch effectively determines when to collaborate and who to collaborate with, balancing information gain and transmission cost. As a result, the proposed MC-DSA achieves better segmentation performance and strong robustness under different wireless channel conditions.

	\subsubsection{Analysis of Cooperative Resource}
	
	Fig. \ref{overall_kpi_cum_bte} illustrates the system efficiency of the proposed MC-DSA under the Rayleigh fading channel in terms of bandwidth, cumulative latency, and cumulative energy.
	For bandwidth, the required bandwidth remains around 1.8 MHz across all 1000 frames, while the allocated bandwidth is kept at 10 MHz. The large bandwidth margin ensures that the proposed MC-DSA can stably support the feature transmission even under fading, without causing congestion or bandwidth shortage. This demonstrates that the communication policy efficiently utilizes the available spectrum and maintains a sufficient safety margin for dynamic channel variation.
	For cumulative latency, the total latency increases steadily with the number of frames, reaching about 33 s after 1000 frames. Within this total, computation latency contributes about 16 s and transmission latency about 17 s, showing a balanced allocation between local processing and communication. The near-linear growth indicates that the proposed MC-DSA maintains stable real-time performance under Rayleigh fading and effectively avoids delay accumulation that would otherwise occur in traditional cooperative schemes.
	For cumulative energy, after 1000 frames, the total energy reaches about 600 J, composed of roughly 240 J from computation and 360 J from communication. The linear increase of both components confirms consistent power allocation, while the smaller slope of computation energy indicates that the proposed MC-DSA dynamically optimizes local inference effort according to the communication state.
	Overall, these results verify that the proposed MC-DSA achieves efficient bandwidth utilization, low latency accumulation, and balanced energy consumption. Its adaptive cooperation mechanism ensures stable system operation and high efficiency in multi-UAV cooperative communication.
	
	\subsubsection{Analysis of Dynamic Environmental Adaptability}
	
	This experiment aims to analyze the dynamic environmental adaptability and verify the semantic transmission effectiveness of the proposed MC-DSA. Fig. \ref{channelsmetric} presents the segmentation performance of communication systems under wireless channels.
	
	Under the AWGN channel, as shown in Fig. \ref{Bench_AWGN}, the proposed MC-DSA exhibits superior results. At an SNR of -10 dB, the proposed MC-DSA reaches about 57\% OAcc, 30\% mAcc, 40\% FWIoU, and 20\% mIoU. When the SNR rises to 0 dB, the proposed MC-DSA improves to 83\% OAcc, 59\% mAcc, 72\% FWIoU, and 50\% mIoU, outperforming ResNet and ECANet by 3\%–5\%. At high SNRs of 10 dB and 20 dB, the proposed MC-DSA maintains around 90\% OAcc and 60\% mIoU, keeping the leading margin. These results demonstrate that the proposed MC-DSA effectively suppresses noise and preserves semantic information through its strategy. As SNR increases, the channel ceases to be the bottleneck, so all methods saturate and the gap shrinks while the proposed MC-DSA remains slightly ahead. The advantage is most visible at low SNR because the proposed MC-DSA transmits and fuses semantically relevant features, which suppresses noise sensitivity and preserves task-critical information.
	
	Under the Rician fading channel, as shown in Fig. \ref{Bench_Rician}, the proposed MC-DSA delivers strong adaptability and a clearer advantage than under the AWGN channel. At -10 dB, the proposed MC-DSA attains 64\% OAcc, 37\% mAcc, 49\% FWIoU, and 27\% mIoU, which is about 2-3\% higher than the benchmarks. At 0 dB, the scores increase to 80\% OAcc, 56\% mAcc, 67\% FWIoU, and 46\% mIoU, preserving a margin of about 2-4\%. At 10 dB and 20 dB, the proposed MC-DSA reaches 89\% OAcc and 67\% mAcc with 80\% FWIoU and 59\% mIoU, keeping a steady lead. 
	The Rician fading channel includes a dominant line of sight component, which preserves semantic feature transmission more effectively. This makes cooperative information exchange more reliable and allows the proposed MC-DSA to better exploit complementary observations from selected S-UAVs. By focusing on semantically relevant S-UAVs and fusing their features with the local T-UAV feature, MC-DSA improves the fused representation and thus achieves better segmentation performance.

	Under the Rayleigh fading channel, as shown in Fig. \ref{Bench_Rayleigh}, there is no dominant line of sight component and deep fades occur, so all methods degrade, yet the proposed MC-DSA remains the best performer. At SNR of -10 dB, the proposed MC-DSA attains about 64\% OAcc, 36\% mAcc, 49\% FWIoU, and 27\% mIoU, ahead of the benchmarks. At 0 dB the scores rise to about 79\% OAcc, 53\% mAcc, 66\% FWIoU, and 42\% mIoU, and at 20 dB the proposed MC-DSA reaches about 88\% OAcc, 65\% mAcc, 79\% FWIoU, and 57\% mIoU, keeping a lead of about 3-4\%. 
	The superiority is less pronounced than in the Rician fading channel because the absence of a stable line of sight path reduces the reliability of cooperative transmission. However, the advantage is more evident than under the AWGN channel, since multipath fading imposes a more challenging transmission condition. In this case, the adaptive semantic selection and cooperative fusion of the proposed MC-DSA help retain more task-relevant information than the benchmarks, thereby yielding a larger performance gap. These results further confirm that the proposed MC-DSA maintains strong robustness across different wireless communication environments.

	
	As shown in Fig. \ref{channelsselacc}, the curves report the selection accuracy in terms of when2com and who2com. All methods improve steadily with increasing SNR and approach near-perfect accuracy at high SNR. For example, at -10 dB under the Rayleigh fading channel, the proposed MC-DSA achieves about 68\% when2com accuracy and 42\% who2com accuracy, which are slightly higher than those of the benchmarks. However, as SNR increases, the performance gap rapidly narrows and all methods converge to over 95\% accuracy.	
	This small gap is expected because when2com and who2com only evaluate the correctness of the collaboration decision, which is a simpler intermediate task than downstream semantic segmentation. Once the channel quality becomes moderate, the transmitted query and key features are already sufficient to support reliable similarity estimation, so the decision accuracy of all methods quickly saturates. 
	The main advantage of the proposed MC-DSA is therefore reflected not only in accurate collaboration decisions, but also in how these decisions improve the downstream segmentation task. By selecting semantically relevant collaborators and enabling effective cooperative fusion, MC-DSA translates similar selection accuracy into better task-level performance under challenging channel conditions.
	
	\subsubsection{Impact of Collaboration Scale and Source Selection}

\begin{figure*}[t]
		\centering
		\includegraphics[width=0.99\textwidth]{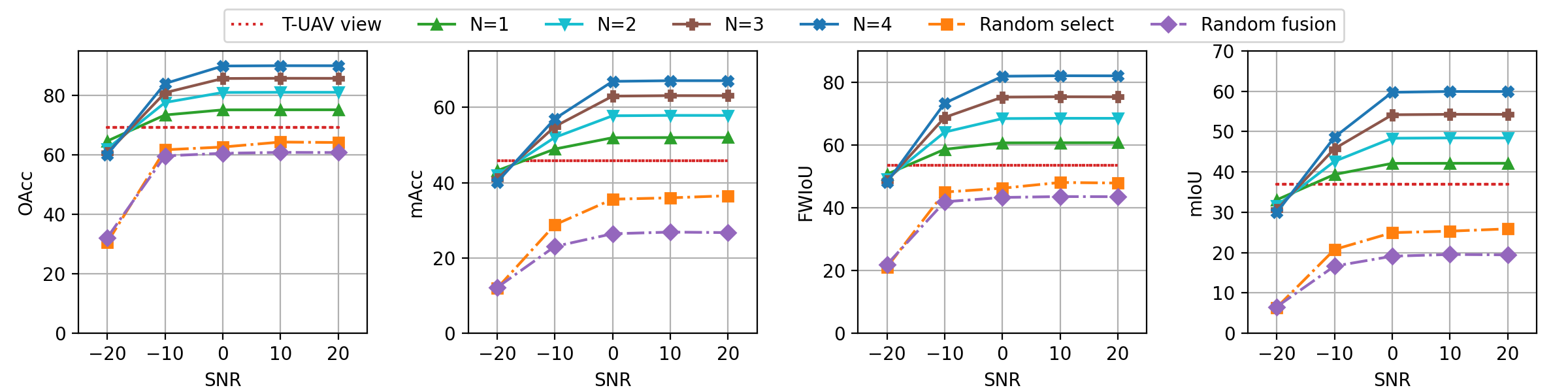}
		\caption{Segmentation performance of the proposed MC-DSA under different collaboration scales and source selection strategies.}
		\label{mode_compare_metrics_AWGN_20}
	\end{figure*}

	As shown in Fig. \ref{mode_compare_metrics_AWGN_20}, we further evaluate the proposed MC-DSA under the AWGN channel to investigate the impact of collaboration scale and source selection.
	
	The T-UAV view corresponds to single-view task execution without cooperation, and thus its performance remains nearly unchanged over the whole SNR range. After introducing S-UAVs, all collaborative settings of the proposed MC-DSA consistently outperform the T-UAV view, confirming the effectiveness of multi-UAV cooperation. Moreover, the performance improves steadily as \(N\) increases from 1 to 4, especially in the high-SNR region. At 20 dB, the proposed MC-DSA with \(N=4\) achieves about 89\% OAcc, 67\% mAcc, 82\% FWIoU, and 60\% mIoU, whereas the \(N=1\) setting attains about 75\% OAcc, 52\% mAcc, 60\% FWIoU, and 42\% mIoU, with the results for \(N=2\) and \(N=3\) falling in between. 
	This is because a larger collaboration scale provides richer complementary observations, which helps recover missing scene semantics and improves the quality of the fused representation for the downstream task.
	
	The comparison with random select and random fusion further reveals the source of the performance gain.  Although these benchmarks also exploit multi-UAV observations, their performance remains consistently inferior to the T-UAV view. At high SNRs, random select reaches only about 64\% OAcc, 36\% mAcc, 48\% FWIoU, and 25\% mIoU, while random fusion performs even worse in mAcc and mIoU. In contrast, the proposed MC-DSA with any collaboration scale maintains a clear advantage. 
	These results indicate that the performance gain comes not simply from involving more UAV views, but from selecting semantically relevant collaborators and performing effective feature aggregation. Otherwise, irrelevant or redundant information may be introduced and even degrade the segmentation performance.

	\subsubsection{Impact of Threshold}

\begin{figure*}[t]
	\centering
	\includegraphics[width=0.99\textwidth]{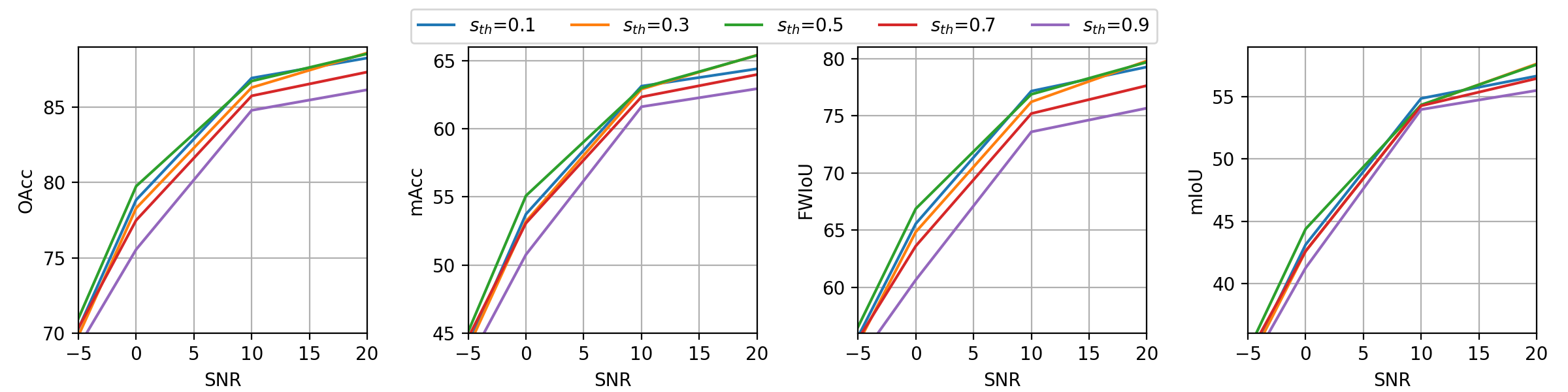}
	\caption{Segmentation performance of the proposed MC-DSA under different thresholds $s_{\mathrm{th}}$.}
	\label{metric_all_sth}
\end{figure*}
	
	\begin{figure*}
		\centering
		\includegraphics[width=0.99\textwidth]{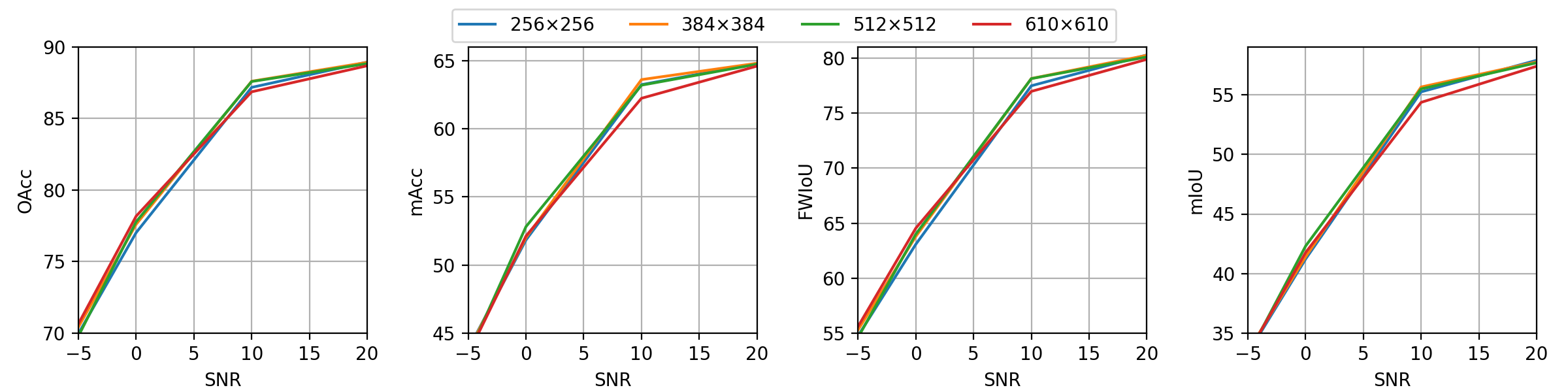}
		\caption{Segmentation performance of the proposed MC-DSA for different input resolutions.}
		\label{metric_all_fenbianlv}
	\end{figure*}
	
	As shown in Fig. \ref{metric_all_sth}, the threshold $s_{\rm{th}}$ has a clear influence on segmentation performance under the Rayleigh fading channel. 
	During training, the model adopts soft fusion, while $s_{\rm{th}}$ is introduced only during inference for source selection.
	When the threshold is high, with $s_{\rm{th}}$ set to 0.9, the model consistently achieves the lowest performance across SNRs. As $s_{\rm{th}}$ decreases from 0.9 to 0.7 and then to 0.5, all metrics improve noticeably. At 0 dB, with $s_{\rm{th}}$ set to 0.9, the proposed MC-DSA yields about 76\% OAcc, 51\% mAcc, 61\% FWIoU, and 41\% mIoU, while lowering the threshold to $s_{\rm{th}}$ set to 0.5 increases them to about 80\% OAcc, 55\% mAcc, 67\% FWIoU, and 44\% mIoU. At 20 dB, with $s_{\rm{th}}$ set to 0.5, MC-DSA achieves about 89\% OAcc, 65\% mAcc, 80\% FWIoU, and 56\% mIoU, whereas with $s_{\rm{th}}$ set to 0.9 it drops to about 86\% OAcc, 63\% mAcc, 76\% FWIoU, and 55\% mIoU. In contrast, further decreasing the threshold from 0.5 to 0.3 or 0.1 brings only marginal changes, and the curves for $s_{\rm{th}}$ set to 0.1, 0.3, and 0.5 are almost overlapped, indicating that performance has already saturated.	
	Overall, $s_{\rm{th}}$ controls how strictly an S-UAV is selected to upload its semantic features. A high threshold is overly conservative and suppresses potentially useful S-UAVs, leading to insufficient complementary multi-view information and thus inferior fusion quality. 
	Lowering the threshold allows more relevant S-UAV features to participate in cooperative fusion, which improves semantic completeness and segmentation accuracy. Once $s_{\rm{th}}$ is reduced to around 0.5, the selection becomes sufficiently inclusive, so further reductions to 0.3 and 0.1 provide limited additional benefit.
	
	\subsubsection{Impact of Resolution}

	As shown in Fig. \ref{metric_all_fenbianlv}, we evaluate the proposed MC-DSA under the Rician fading channel using different resolutions. The curves are almost overlapped, which indicates that the proposed cooperative strategy generalizes well to input-scale changes. At -5 dB, all resolutions yield very close results, with OAcc around 70\%, mAcc around 44\%, FWIoU around 55\%, and mIoU around 33\%, and the resolution-induced gaps are visually negligible. At 20 dB, the performance almost coincides across resolutions, reaching about 89\% OAcc, 65\% mAcc, 80\% FWIoU, and 58\% mIoU, with only marginal differences among resolutions.	
	These results suggest that the proposed MC-DSA maintains stable segmentation quality under varying sensing scales, and that the policy decision process remains robust to such input changes. This is mainly because adaptive pooling is used during query and key generation to normalize feature maps to a fixed dimension, making the similarity computation largely insensitive to resolution. Therefore, the proposed MC-DSA can accommodate different onboard compute budgets while preserving the final task performance.

	\subsubsection{Sensitivity to Learning Rate}
	
	\begin{figure*}
		\centering
		\includegraphics[width=0.99\textwidth]{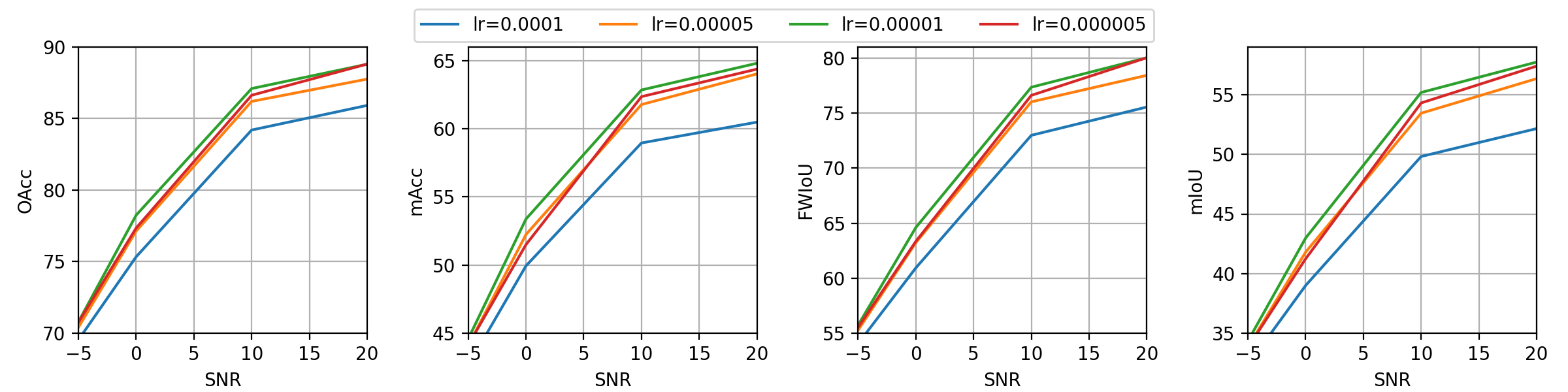}
		\caption{Segmentation performance comparison of the proposed MC-DSA under different learning rates.}
		\label{metric_lr_overall}
	\end{figure*}

	As shown in Fig. \ref{metric_lr_overall}, the experiment compares the performance of the proposed MC-DSA under the Rician fading channel using different learning rates. The model trained with 0.00001 achieves the best performance. 0.000005 is a close second, followed by 0.00005, while 0.0001 yields the weakest results. At 10 dB, using 0.00001 gives 87\% OAcc, 63\% mAcc, 77\% FWIoU, and 55\% mIoU. At 20 dB, 0.00001 reaches 89\% OAcc, 65\% mAcc, 80\% FWIoU, and 58\% mIoU. It indicates that decreasing the learning rate from 0.0001 to 0.00001 brings clear and consistent gains, while reducing it further to 0.000005 may slightly weaken metrics. 
	This reflects the training dynamics. A large learning rate causes unstable parameter updates, leading to slower convergence and suboptimal feature alignment. Conversely, a small rate provides smooth updates but limits adaptation speed. The intermediate rate 0.00001 strikes the best balance between stability and convergence efficiency, enabling more accurate learning and more effective cooperation.

	\subsubsection{Training Convergence with Adaptive Learning Rate}
	
	\begin{figure}[t]
		\centering
		\subfloat[Training loss\label{train_loss}]%
		{\includegraphics[width=0.48\linewidth]{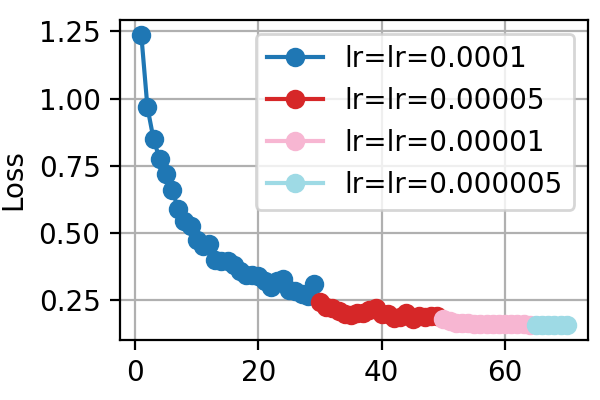}}\hfill
		\subfloat[Validation OAcc\label{val_overall}]%
		{\includegraphics[width=0.48\linewidth]{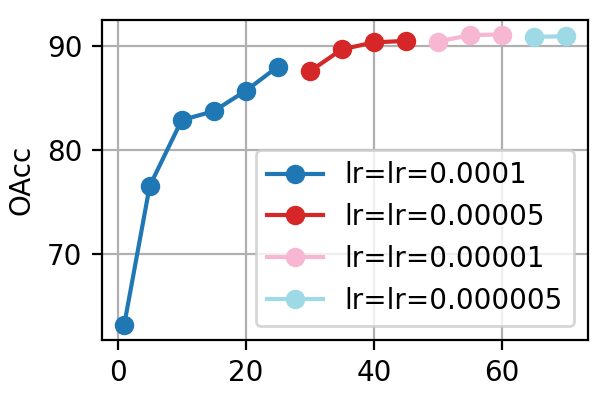}}
		\caption{Training loss and validation OAcc of the proposed MC-DSA under the AWGN channel.}
	\end{figure}
	
	In Fig. \ref{train_loss} and Fig. \ref{val_overall}, the proposed MC-DSA converges smoothly under an adaptive learning-rate schedule. With lr is 0.0001, the training loss drops rapidly from 1.24 to 0.29, and the validation OAcc rises from 63.3\% at epoch 0 to 88.0\% at epoch 25. After the learning rate decays to 0.00005, the loss decreases further to 0.19 and the validation OAcc increases to 90.4\% at epoch 45. After the learning rate decays to 0.00001, the loss is reduced to 0.16 and the validation OAcc reaches 91.0\% at epoch 60. With lr is 0.000005, the loss stabilizes at 0.155 and the validation OAcc remains at 91.0\% at epoch 70. 
	This reflects the trade-off between exploration and stability. A larger rate speeds early progress but leaves a higher loss floor because semantic encoder updates and policy gating are aggressive. A small rate yields stable optimization but slow improvement. Intermediate rates around 0.00001 provide the best balance. Decays of learning rate achieve fast convergence and strong final accuracy. Very small additional gains after the last decay suggest that training can be stopped once the loss flattens and validation accuracy saturates.
	
	\section{Conclusion}
	\label{Conclusion}
	
	This paper has proposed the MC-DSA, which deeply integrates DL technology with UAV cooperative mechanisms to enhance the efficiency task execution. Simulation experiments demonstrate that under low-quality images and dynamic environments, the proposed MC-DSA network outperforms benchmarks, particularly achieving improvements in the segmentation performance metrics. Experimental data results validate the MC-DSA's effectiveness in different scenarios while revealing its strong adaptability and robustness to dynamic environmental changes. These characteristics hold significant value for UAV applications in the low-altitude economy, enabling effective responses to rapidly changing conditions and real-time requirements. Future work will focus on algorithmic lightweight design and real-world scenario expansion.
	
    \vspace{-1 mm} 
	
	\bibliographystyle{IEEEtrans}   
	\bibliography{Ref-FigureSemantic}

@article{li2020energy,
  title={Energy-efficient UAV-assisted mobile edge computing: Resource allocation and trajectory optimization},
  author={Li, Mushu and Cheng, Nan and Gao, Jie and Wang, Yinlu and Zhao, Lian and Shen, Xuemin},
  journal={IEEE Transactions on Vehicular Technology},
  volume={69},
  number={3},
  pages={3424--3438},
  year={2020},
  publisher={IEEE}
}

@inproceedings{guan2021energy,
  title={Energy-efficient uav communication with 3d trajectory optimization},
  author={Guan, Zixuan and Wang, Siye and Gao, Luoyu and Xu, Wenbo},
  booktitle={2021 7th International Conference on Computer and Communications (ICCC)},
  pages={312--317},
  year={2021},
  organization={IEEE}
}

@inproceedings{liu2021swin,
  title={Swin transformer: Hierarchical vision transformer using shifted windows},
  author={Liu, Ze and Lin, Yutong and Cao, Yue and Hu, Han and Wei, Yixuan and Zhang, Zheng and Lin, Stephen and Guo, Baining},
  booktitle={Proceedings of the IEEE/CVF international conference on computer vision},
  pages={10012--10022},
  year={2021}
}

@ARTICLE{mUAV1,
  author={Xu, Yu and Zhang, Tiankui and Liu, Yuanwei and Yang, Dingcheng and Xiao, Lin and Tao, Meixia},
  journal={IEEE Transactions on Wireless Communications}, 
  title={UAV-Assisted MEC Networks With Aerial and Ground Cooperation}, 
  year={2021},
  volume={20},
  number={12},
  pages={7712-7727},
  doi={10.1109/TWC.2021.3086521}}

@ARTICLE{segmentpaper,
  author={Wang, Qi and Jing, Wei and Chi, Kaichen and Yuan, Yuan},
  journal={IEEE Transactions on Geoscience and Remote Sensing}, 
  title={Cross-Difference Semantic Consistency Network for Semantic Change Detection}, 
  year={2024},
  volume={62},
  number={},
  pages={1-12},
  doi={10.1109/TGRS.2024.3386334}}

@ARTICLE{mIOU,
  author={Mei, Liye and Ye, Zhaoyi and Xu, Chuan and Wang, Hongzhu and Wang, Ying and Lei, Cheng and Yang, Wei and Li, Yansheng},
  journal={IEEE Transactions on Geoscience and Remote Sensing}, 
  title={SCD-SAM: Adapting Segment Anything Model for Semantic Change Detection in Remote Sensing Imagery}, 
  year={2024},
  volume={62},
  number={},
  pages={1-13},
  doi={10.1109/TGRS.2024.3407884}}

@ARTICLE{airsim,
  author={Pan, Tingwei and Dong, Hongbin and Deng, Baosong and Gui, Jianjun and Zhao, Bingxu},
  journal={IEEE Signal Processing Letters}, 
  title={Robust Cross-Drone Multi-Target Association Using 3D Spatial Consistency}, 
  year={2024},
  volume={31},
  number={},
  pages={71-75},
  doi={10.1109/LSP.2023.3341009}}

@INPROCEEDINGS{ResNet,
  author={He, Kaiming and Zhang, Xiangyu and Ren, Shaoqing and Sun, Jian},
  booktitle={2016 IEEE Conference on Computer Vision and Pattern Recognition (CVPR)}, 
  title={Deep Residual Learning for Image Recognition}, 
  year={2016},
  volume={},
  number={},
  pages={770-778},
  doi={10.1109/CVPR.2016.90}}

@INPROCEEDINGS{ECA–Net,
  author={Wang, Qilong and Wu, Banggu and Zhu, Pengfei and Li, Peihua and Zuo, Wangmeng and Hu, Qinghua},
  booktitle={2020 IEEE/CVF Conference on Computer Vision and Pattern Recognition (CVPR)}, 
  title={ECA-Net: Efficient Channel Attention for Deep Convolutional Neural Networks}, 
  year={2020},
  volume={},
  number={},
  pages={11531-11539},
  doi={10.1109/CVPR42600.2020.01155}}

@ARTICLE{SegNet,
  author={Badrinarayanan, Vijay and Kendall, Alex and Cipolla, Roberto},
  journal={IEEE Transactions on Pattern Analysis and Machine Intelligence}, 
  title={SegNet: A Deep Convolutional Encoder-Decoder Architecture for Image Segmentation}, 
  year={2017},
  volume={39},
  number={12},
  pages={2481-2495},
  doi={10.1109/TPAMI.2016.2644615}}

@ARTICLE{tiaozhan,
  author={Yang, Wanting and Du, Hongyang and Liew, Zi Qin and Lim, Wei Yang Bryan and Xiong, Zehui and Niyato, Dusit and Chi, Xuefen and Shen, Xuemin and Miao, Chunyan},
  journal={IEEE Communications Surveys \& Tutorials}, 
  title={Semantic Communications for Future Internet: Fundamentals, Applications, and Challenges}, 
  year={2023},
  volume={25},
  number={1},
  pages={213-250},
  doi={10.1109/COMST.2022.3223224}
  }

@article{zhishiku,
  title={6G networks: Beyond Shannon towards semantic and goal-oriented communications},
  author={Strinati, Emilio Calvanese and Barbarossa, Sergio},
  journal={Computer Networks},
  volume={190},
  pages={107930},
  year={2021},
  publisher={Elsevier}
}

@ARTICLE{engineeringsemantic,
  author={Wheeler, Dylan and Natarajan, Balasubramaniam},
  journal={IEEE Access}, 
  title={Engineering Semantic Communication: A Survey}, 
  year={2023},
  volume={11},
  number={},
  pages={13965-13995},
  doi={10.1109/ACCESS.2023.3243065}}

@ARTICLE{yuyichaoyuebite,
  author={Gündüz, Deniz and Qin, Zhijin and Aguerri, Inaki Estella and Dhillon, Harpreet S. and Yang, Zhaohui and Yener, Aylin and Wong, Kai Kit and Chae, Chan-Byoung},
  journal={IEEE Journal on Selected Areas in Communications}, 
  title={Beyond Transmitting Bits: Context, Semantics, and Task-Oriented Communications}, 
  year={2023},
  volume={41},
  number={1},
  pages={5-41},
  doi={10.1109/JSAC.2022.3223408}}

@ARTICLE{crossmodal,
  author={Li, Ang and Wei, Xin and Wu, Dan and Zhou, Liang},
  journal={IEEE Wireless Communications}, 
  title={Cross-Modal Semantic Communications}, 
  year={2022},
  volume={29},
  number={6},
  pages={144-151},
  doi={10.1109/MWC.008.2200180}}

@ARTICLE{yuyiganzhi,
  author={Shi, Guangming and Xiao, Yong and Li, Yingyu and Xie, Xuemei},
  journal={IEEE Communications Magazine}, 
  title={From Semantic Communication to Semantic-Aware Networking: Model, Architecture, and Open Problems}, 
  year={2021},
  volume={59},
  number={8},
  pages={44-50},
  doi={10.1109/MCOM.001.2001239}
  }

@ARTICLE{multiuser1,
  author={Huang, Wei and Wang, Jun and Chen, Xiaonan and Peng, Qihang and Zhu, Yichao},
  journal={IEEE Communications Letters}, 
  title={Flag Vector Assisted Multi-User Semantic Communications for Downlink Text Transmission}, 
  year={2024},
  volume={28},
  number={6},
  pages={1283-1287},
  doi={10.1109/LCOMM.2024.3389709}}

@ARTICLE{multiuser2,
  author={Xie, Huiqiang and Qin, Zhijin and Li, Geoffrey Ye},
  journal={IEEE Wireless Communications Letters}, 
  title={Task-Oriented Multi-User Semantic Communications for VQA}, 
  year={2022},
  volume={11},
  number={3},
  pages={553-557},
  doi={10.1109/LWC.2021.3136045}}

@INPROCEEDINGS{multiuser3,
  author={Chen, Weixuan and Yang, Qianqian and Yang, Zhaohui and Duan, Yiping and Zhang, Zhaoyang},
  booktitle={2024 IEEE International Conference on Image Processing (ICIP)}, 
  title={Pilot-Free Semantic Communication Over Multi-User Mimo Fading Channels}, 
  year={2024},
  volume={},
  number={},
  pages={3958-3964},
  doi={10.1109/ICIP51287.2024.10648086}}

@INPROCEEDINGS{multiuser4,
  author={Wei, Shuangying and Feng, Chunyan and Guo, Caili and Zhang, Biling and Chen, Jiujiu},
  booktitle={2024 IEEE/CIC International Conference on Communications in China (ICCC Workshops)}, 
  title={Deep Over-the-Air Computation Based Multi-User Semantic Communication for Smart Cities}, 
  year={2024},
  volume={},
  number={},
  pages={622-626},
  doi={10.1109/ICCCWorkshops62562.2024.10693735}}

@INPROCEEDINGS{multiuser5,
  author={Zeng, Youcheng and He, Xinxin and Chen, Xu and Tong, Haonan and Yang, Zhaohui and Guo, Yijun and Hao, Jianjun},
  booktitle={ICC 2024 - IEEE International Conference on Communications}, 
  title={DMCE: Diffusion Model Channel Enhancer for Multi-User Semantic Communication Systems}, 
  year={2024},
  volume={},
  number={},
  pages={855-860},
  doi={10.1109/ICC51166.2024.10622730}}

@INPROCEEDINGS{multiuser6,
  author={Xing, Huanlai and Zhang, Haolan and Wang, Xinhan and Xu, Lexi and Xiao, Zhiwen and Zhao, Bowen and Luo, Shouxi and Feng, Li and Dai, Yuanshun},
  booktitle={2023 IEEE International Conference on Communications Workshops (ICC Workshops)}, 
  title={A Multi-user Deep Semantic Communication System based on Federated Learning with Dynamic Model Aggregation}, 
  year={2023},
  volume={},
  number={},
  pages={1612-1616},
  doi={10.1109/ICCWorkshops57953.2023.10283725}}

@ARTICLE{IEEE6GEdge,
  author={Huang, Kaibin and Lan, Qiao and Liu, Zhiyan and Yang, Lin},
  journal={IEEE Communications Magazine}, 
  title={Semantic Data Sourcing for 6G Edge Intelligence}, 
  year={2023},
  volume={61},
  number={12},
  pages={70-76},
  doi={10.1109/MCOM.001.2200962}}

@inproceedings{liu2020when2com,
  title={When2com: Multi-agent perception via communication graph grouping},
  author={Liu, Yen-Cheng and others},
  booktitle={Proceedings of the IEEE/CVF Conference on Computer Vision and Pattern Recognition},
  year={2020}
}

@ARTICLE{maritimeUAV,
  author={Nomikos, Nikolaos and Gkonis, Panagiotis K. and Bithas, Petros S. and Trakadas, Panagiotis},
  journal={IEEE Open Journal of the Communications Society}, 
  title={A Survey on UAV-Aided Maritime Communications: Deployment Considerations, Applications, and Future Challenges}, 
  year={2023},
  volume={4},
  number={},
  pages={56-78},
  doi={10.1109/OJCOMS.2022.3225590}}

@ARTICLE{yuyituxiang1,
  author={Zhang, Hongwei and Shao, Shuo and Tao, Meixia and Bi, Xiaoyan and Letaief, Khaled B.},
  journal={IEEE Journal on Selected Areas in Communications}, 
  title={Deep Learning-Enabled Semantic Communication Systems With Task-Unaware Transmitter and Dynamic Data}, 
  year={2023},
  volume={41},
  number={1},
  pages={170-185},
  doi={10.1109/JSAC.2022.3221991}
  }
	
\newpage

\begin{IEEEbiography}[{\includegraphics[width=1.1in,height=1.4in,clip,keepaspectratio]{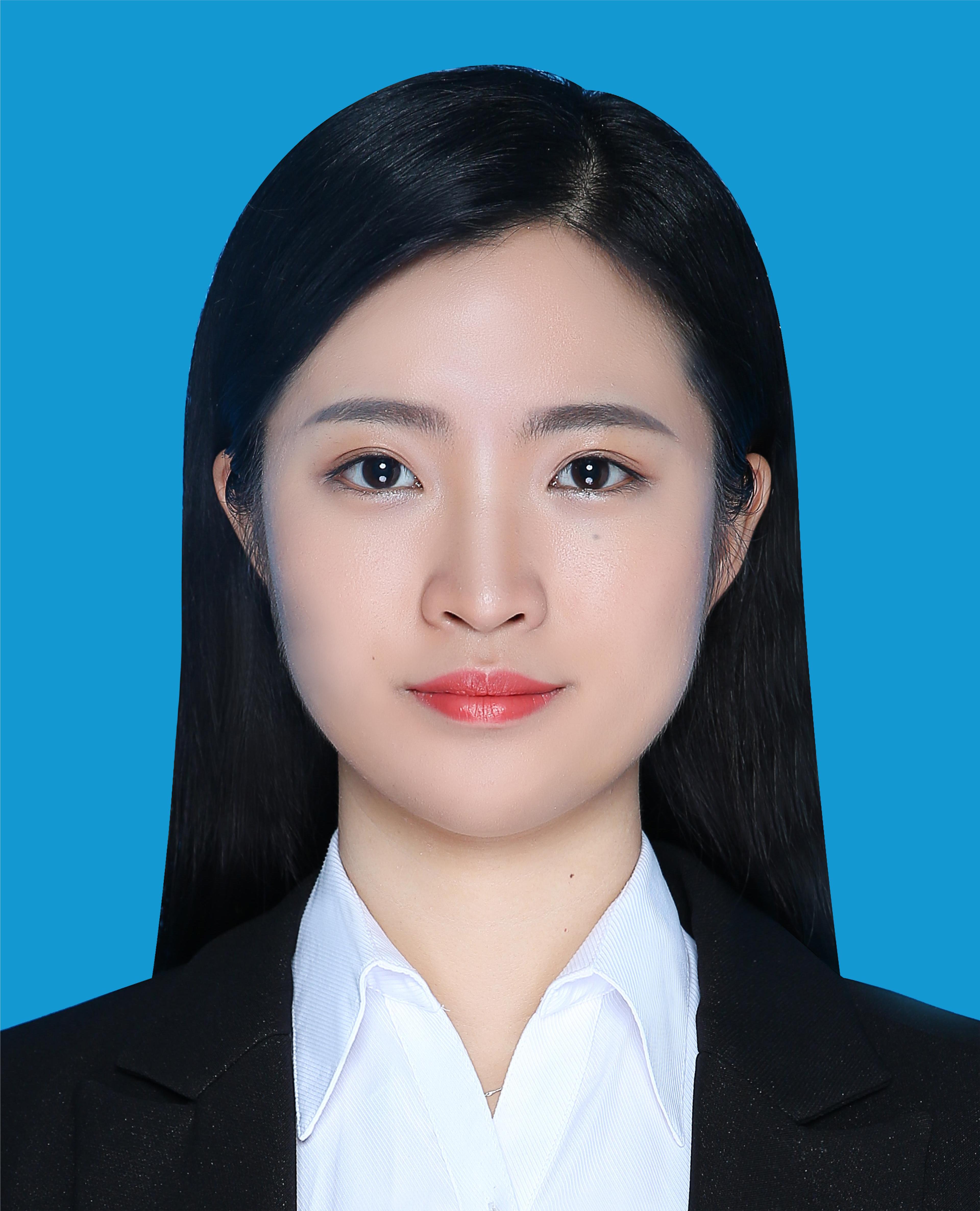}}]{Xiaoling Han}
	(Student Member, IEEE) received the B.S. degree in Electronic Information Engineering from Dalian Maritime University, Dalian, China, in 2020. She is currently working toward the Ph.D. degree at the Department of Information Science and Technology, Dalian Maritime University, Dalian, China, under the supervision of Prof. B. Lin. Her research interests include maritime communications, UAV-based communications and semantic communications.
\end{IEEEbiography}
\vspace{-14 mm} 
\begin{IEEEbiography}[{\includegraphics[width=1.1in,height=1.4in,clip,keepaspectratio]{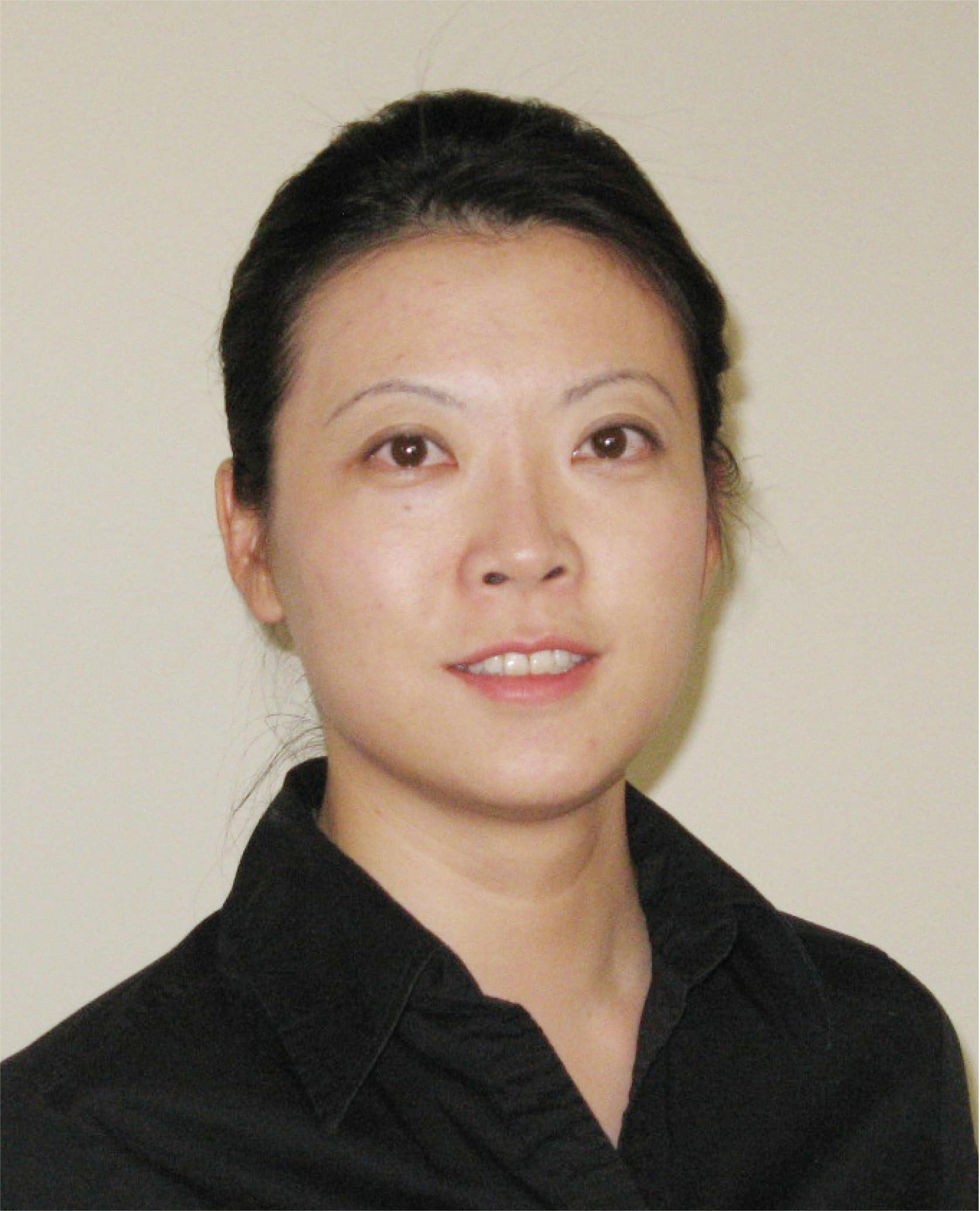}}]{Bin Lin}
	
	(Senior Member, IEEE) received the B.S. and M.S. degrees from Dalian Maritime University, Dalian, China, in 1999 and 2003 respectively, and the Ph.D. degree from the Broadband Communications Research Group, Department of Electrical and Computer Engineering, University of Waterloo, Waterloo, ON, Canada, in 2009. She is currently a Full Professor and Dean of Communication Engineering Department, School of Information Science and Technology, Dalian Maritime University. She has been a Visiting Scholar with George Washington University, Washington, DC, USA, from 2015 to 2016. Her current research interests include wireless communications, network dimensioning and optimization, resource allocation, artificial intelligence, maritime communication networks, edge/cloud computing, wireless sensor networks, and Internet of Things. Dr. Lin is an associate editor of IEEE Transactions on Vehicular Technology and IEEE Internet of Things Journal.
	
\end{IEEEbiography}
\vspace{-16 mm}
\begin{IEEEbiography}[{\includegraphics[width=1.1in,height=1.4in,clip,keepaspectratio]{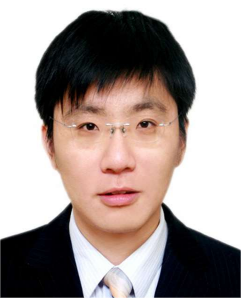}}]{Nan Wu}
	
	(Member, IEEE) received the B.S. degree in electrical and electronics engineering from Dalian University of Technology, Dalian, China, in 2003 and the M.Sc. and Ph.D. degrees from the University of Southampton, Southampton, U.K., in 2004 and 2008, respectively. From 2008 to 2009, he was a Guest Researcher with the National Institute of Standards and Technology, Gaithersburg, MD, USA. Since 2010, he has been an Associate Professor with the School of Electrical and Information Engineering, Dalian Maritime University, Dalian. He has authored one book, more than 50 articles. He also holds two patents. His research interests include deep-learning-based communications, visible light communication, indoor positioning and MIMO.
	
\end{IEEEbiography}
\vspace{-15 mm}
\begin{IEEEbiography}[{\includegraphics[width=1.1in,height=1.4in,clip,keepaspectratio]{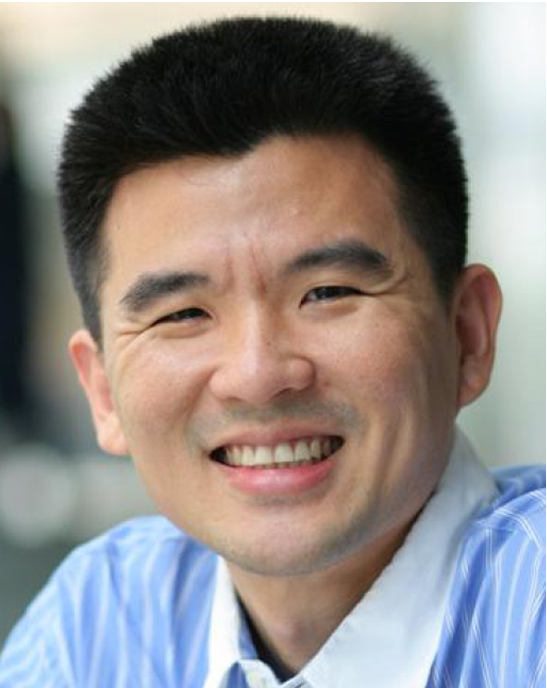}}]{Tony Q. S. Quek}
	
	(S'98-M'08-SM'12-F'18) received the B.E.\ and M.E.\ degrees in electrical and electronics engineering from the Tokyo Institute of Technology in 1998 and 2000, respectively, and the Ph.D.\ degree in electrical engineering and computer science from the Massachusetts Institute of Technology in 2008. Currently, he is the Associate Provost (AI \& Digital Innovation) and Cheng Tsang Man Chair Professor with Singapore University of Technology and Design (SUTD). He also serves as the Director of the Future Communications R\&D Programme, and the ST Engineering Distinguished Professor. He is a co-founder of Silence Laboratories and NeuroRAN. His current research topics include wireless communications and networking, network intelligence, non-terrestrial networks, open radio access network, AI-RAN, and 6G.
	
	Dr.\ Quek was honored with the 2008 Philip Yeo Prize for Outstanding Achievement in Research, the 2012 IEEE William R. Bennett Prize, the 2015 SUTD Outstanding Education Awards -- Excellence in Research, the 2016 IEEE Signal Processing Society Young Author Best Paper Award, the 2017 CTTC Early Achievement Award, the 2017 IEEE ComSoc AP Outstanding Paper Award, the 2020 IEEE Communications Society Young Author Best Paper Award, the 2020 IEEE Stephen O. Rice Prize, the 2020 Nokia Visiting Professor, the 2022 IEEE Signal Processing Society Best Paper Award, the 2024 IIT Bombay International Award For Excellence in Research in Engineering and Technology, the IEEE Communications Society WTC Recognition Award 2024, and the Public Administration Medal (Bronze). He is an IEEE Fellow, a WWRF Fellow, an AIIA Fellow, and a Fellow of the Academy of Engineering Singapore.
\end{IEEEbiography}

\end{document}